\documentclass[a4paper,11pt]{article}
\pdfoutput=1 

\usepackage{jheppub} 

\usepackage[T1]{fontenc} 

\title{\boldmath Probing the Energy Evolution and Hadronic Cascade Time Dependence of Forward-Backward Multiplicity Correlations in pp Collisions at $\sqrt{s} = 0.9$--$13.6$~TeV with AMPT}

\author{Dibakar Dhar,}
\author{Tumpa Biswas,}
\author[1]{Prabir Kumar Haldar\note{Corresponding author.}}

\affiliation{Department of Physics, Cooch Behar Panchanan Barma University, \\Vivekananda Street, Coochbehar 736101, India}

\emailAdd{dibakar.dhar@cern.ch}
\emailAdd{tumpa.biswas@cern.ch}
\emailAdd{prabir.kumar.haldar@cern.ch}

\abstract{Forward-backward (FB) multiplicity correlations are studied in proton--proton collisions at centre-of-mass energies of $\sqrt{s} = 0.9$, $2.76$, $5.02$, $7$, $13$, and $13.6$~TeV using the AMPT model in its default (DF) and string melting (SM) configurations. The coefficient $b_{\rm corr}$ is obtained for windows separated in pseudorapidity and azimuth, and its variation with the pseudorapidity gap, window width, collision energy, transverse momentum interval, and hadronic cascade time $\tau_{\rm HC}$ is investigated. The simulated results are compared with the available ALICE measurements, and the model is validated against experimental pseudorapidity densities and charged-pion transverse momentum spectra. The correlation strength is found to decrease gradually with increasing pseudorapidity gap while retaining a substantial pedestal, to rise non-linearly with the window width, and to increase substantially with collision energy. The DF configuration yields systematically larger values of $b_{\rm corr}$ than the SM configuration, while the SM configuration provides the better description of the ALICE data at $0.9$ and $2.76$~TeV and the DF configuration at $7$~TeV. In the lowest transverse momentum intervals the deviations from the data are largest for both configurations. The hadronic cascade time is varied within the SM configuration over $\tau_{\rm HC} = 0.6$, $5$, $15$, and $30$~fm/$c$. A monotonic reduction of $b_{\rm corr}$ with increasing $\tau_{\rm HC}$ is observed at all energies and for all window widths, with the ordering of the curves preserved throughout. The effect is modest but statistically significant. These results indicate that the FB correlation strength is primarily imprinted at the partonic stage and only mildly diluted by the subsequent hadronic evolution.}

\begin{document} 
\maketitle
\flushbottom

\section{Introduction}
\label{sec:intro}

Ultrarelativistic hadronic collisions have been studied intensively for several decades, and this effort has substantially deepened our picture of the strong interaction and of how nuclear matter behaves in extreme conditions. Among the observables that drive this programme, the multiplicity of charged particles plays a central role, since its distribution over the full phase space or over restricted regions encodes information about the processes that produce particles. These processes are shaped both by the initial state of the collision and by its later dynamical evolution, which is why multiplicity measurements have remained a major focus of experimental and theoretical work alike. Experimental programmes at the Relativistic Heavy Ion Collider (RHIC), located at Brookhaven National Laboratory, and at the Large Hadron Collider (LHC) at CERN have generated extensive data sets, and a range of phenomenological models has been developed alongside them to interpret the observations.

Fluctuation and correlation observables have long played a central role in heavy-ion physics, largely because of their potential to expose the dynamics of particle production. Among the various correlation techniques that have been developed, the forward-backward (FB) analysis is one of the earliest to have been applied in high-energy collisions. The strength of the FB correlation is quantified by the coefficient $b_{\rm corr}$, extracted from a linear regression of the mean multiplicity in the backward rapidity hemisphere, $\langle n_{\rm B} \rangle$, against the event multiplicity in the forward hemisphere, $n_{\rm F}$:

\begin{equation}
\langle n_{\rm B} \rangle = a + b_{\rm corr}\, n_{\rm F} \, .
\label{eq:bcorr_def}
\end{equation}

This linear relation, Eq.~\eqref{eq:bcorr_def}, has been confirmed experimentally~\cite{ref1,ref2,ref3,ref4} and is discussed in Refs.~\cite{ref5,ref6,ref7}. Assuming a linear correlation between $n_{\rm F}$ and $n_{\rm B}$, the Pearson correlation coefficient

\begin{equation}
b_{\rm corr} = \frac{\langle n_{\rm F} n_{\rm B} \rangle - \langle n_{\rm F} \rangle \langle n_{\rm B} \rangle}{\langle n_{\rm F}^{2} \rangle - \langle n_{\rm F} \rangle^{2}}
\label{eq:bcorr_pearson}
\end{equation}

can be used to determine $b_{\rm corr}$ experimentally~\cite{ref2}. Since the intercept $a$ is fixed by $a = \langle n_{\rm B} \rangle - b_{\rm corr} \langle n_{\rm F} \rangle$, it carries no additional information and is usually not considered.

Forward-backward multiplicity correlations have so far been investigated experimentally across a wide range of collision systems, including $e^{+}e^{-}$, $\mu^{+}p$, $pp$, $p\bar{p}$, and $A$--$A$ interactions~\cite{ref3,ref4,ref8,ref9,ref10,ref11,ref12,ref13}. In $e^{+}e^{-}$ annihilation at $\sqrt{s} = 29$ GeV, no such correlations were seen. This null result was attributed to independent fragmentation of the forward and backward jets produced in the process~\cite{ref14}. In contrast, sizeable positive FB correlations have been reported in $pp$ collisions at the ISR~\cite{ref13} at $\sqrt{s} = 52.6$ GeV~\cite{ref4} and in $p\bar{p}$ interactions at the S$p\bar{p}$S collider~\cite{ref15}. Their strength was found to grow markedly with collision energy~\cite{ref3}, a trend later confirmed at much higher energies ($\sqrt{s} \gtrsim 1$ TeV) in $p\bar{p}$ collisions by the E735 collaboration at the Tevatron~\cite{ref12} and in $pp$ collisions by the ATLAS experiment at the LHC ($\sqrt{s} = 0.9$ and $7$ TeV)~\cite{ref16}. One finding reported by ATLAS is that $b_{\rm corr}$ decreases as the minimum transverse momentum of charged particles increases.

The ALICE collaboration~\cite{alice_main} has examined forward-backward (FB) multiplicity correlations in proton-proton (pp) collisions at $\sqrt{s} = 0.9$, $2.76$, and $7$ TeV. The correlation strength was found to increase substantially with collision energy. This dependence cannot be attributed to the growth of the mean multiplicity within the windows alone. The STAR collaboration at RHIC has analysed FB multiplicity correlations in $pp$ and Au--Au collisions at $\sqrt{s_{\mathrm{NN}}} = 200$ GeV~\cite{ref17}. In Au--Au collisions the correlation proved strong, whereas $b_{\rm corr}$ in $pp$ collisions turned out to be rather small ($\sim 0.1$). We attribute this contrast to the narrower pseudorapidity windows adopted here, compared with those used in earlier $pp$ and $p\bar{p}$ measurements.

These correlations have also attracted considerable theoretical attention in high-energy $pp$ and $A$--$A$ collisions. The Dual Parton Model (DPM)~\cite{ref2} and the Quark Gluon String Model (QGSM)~\cite{ref21,ref22} offered the earliest framework for describing them~\cite{ref7,ref18,ref19,ref20}, since both provide a quantitative account of multiparticle production in soft processes. Later refinements of these models~\cite{ref23,ref24,ref25,ref26} introduced collective effects that originate from interactions among strings --- a feature of particular importance in $A$--$A$ collisions. Such effects are formulated within the String Fusion Model (SFM)~\cite{ref27,ref28}. String interactions of this kind were found to alter the FB correlation strength appreciably, and at the same time to reduce multiplicities, raise the mean particle $p_{\rm T}$, and enhance heavy-flavour production in central $A$--$A$ collisions~\cite{ref23,ref29,ref30}. Alongside these experimental efforts, a large body of simulation-based work has addressed the same observable. S.~Bhattacharyya~\cite{spb} examined FB multiplicity correlations in high-energy nucleus--nucleus interactions at a few $A$~GeV/$c$ within the UrQMD and AMPT frameworks. In a related study, S.~Paul~\cite{s_paul} investigated FB correlations together with multiplicity fluctuations of pions produced in $p$--$p$ collisions at recent LHC energies. A. Ahmed \textit{et al.}~\cite{a_ahmed,sir_1} have also examined FB correlations of pions in ultra-relativistic ring- and jet-like events from $^{16}$O--Ag/Br interactions at $E_{\rm lab} = 60\,A$~GeV.

It is conventional to separate FB correlations into short- and long-range contributions~\cite{ref2,ref7}. Within phenomenological models, short-range correlations (SRC) are taken to be confined to a narrow interval of $\eta$-differences, typically up to one unit. Their origin lies in a variety of short-range mechanisms associated with the fragmentation of a single source, such as particles emerging from cluster or resonance decays, as well as correlations induced by jets and mini-jets. Long-range correlations (LRC), by contrast, span a broader range in $\eta$. They arise from fluctuations in the number and properties of the sources that emit particles --- clusters, cut pomerons, strings, mini-jets, and similar objects~\cite{ref2,ref7,ref19,ref23,ref24,ref25,ref26}.

Within the SFM, the fusion of strings is expected to suppress the variance in their number, which in turn implies a weakening of long-range multiplicity correlations~\cite{ref23,ref25,ref26}. Long-range correlations obtained in the Color Glass Condensate (CGC) framework~\cite{ref31}, however, have been reported to grow with collision centrality~\cite{ref32}, in apparent conflict with the SFM expectation. Because of this tension, measurements of correlations between different observables in two well-separated $\eta$ intervals are regarded as a sensitive probe of the initial conditions of hadronic interactions~\cite{ref33}. For $A$--$A$ collisions, such correlations, extending over a wide $\eta$ range, are anticipated to encode the earliest stage of the collision, with only marginal influence from final-state effects~\cite{ref32,ref34}. Insight into the dynamics of $A$--$A$ collisions can be gained from $pp$ measurements by examining how FB correlations respond to collision energy, pseudorapidity, azimuthal angle, and transverse momentum. The role of the hadronic cascade time in shaping particle production has been explored in Refs.~\cite{pradhan,singha}, where it was shown to have a significant impact on the final-state observables. This consideration motivates the present study of multiplicity correlations in $pp$ collisions across a range of energies and its extension to the hadronic cascade time. In this article, we probe the energy evolution and the dependence on hadronic cascade time of forward-backward multiplicity correlations in $pp$ collisions at $\sqrt{s} = 0.9$--$13.6$~TeV, using the AMPT model.

This article is organised as follows. Section~\ref{sec:data_analysis} describes the simulation setup, the AMPT default and string melting modes, the definition of the counting windows, and the procedure used to extract the correlation coefficient. Section~\ref{sec:mult_sep} presents multiplicity correlations in pseudorapidity-separated windows, covering the dependence on the gap between windows, the window width, and the collision energy. Section~\ref{sec:eta_phi} extends the analysis to windows separated in both pseudorapidity and azimuth. Section~\ref{sec:pt} examines the dependence of the correlations on the transverse momentum interval, and Section~\ref{sec:hadron_cascade_time} addresses the role of the hadronic cascade time. Finally, Section~\ref{sec:conclusions} summarises our findings.

\section{Data Analysis}
\label{sec:data_analysis}

\subsection{Simulation Details}
\label{sub_sec:simulation}

The present analysis is based on simulated events generated with the AMPT (A Multi-Phase Transport) model~\cite{ampt_ref}, which describes the evolution of a heavy-ion or pp collision in several stages: the initial state, partonic scattering, hadronisation, and hadronic rescattering. The model has been employed extensively in the study of high-energy nuclear collisions \cite{ampt_1,ampt_2, ampt_3}. Two configurations of the model are employed in this work, namely the default (DF) mode and the string melting (SM) mode, and they are described separately below.

Unless stated otherwise, charged particles are accepted within the kinematic range $0.3 < p_{\rm T} < 1.5$~GeV/$c$. The sole exception is the study of the transverse-momentum dependence presented in Section~\ref{sec:pt}, for which the range is widened to $0.3 < p_{\rm T} < 6$~GeV/$c$. All particles are required to lie within the pseudorapidity interval $|\eta| < 0.8$, which defines the region over which the counting windows are constructed.

\begin{figure}[!h]
\begin{center}
\includegraphics[scale=0.75]{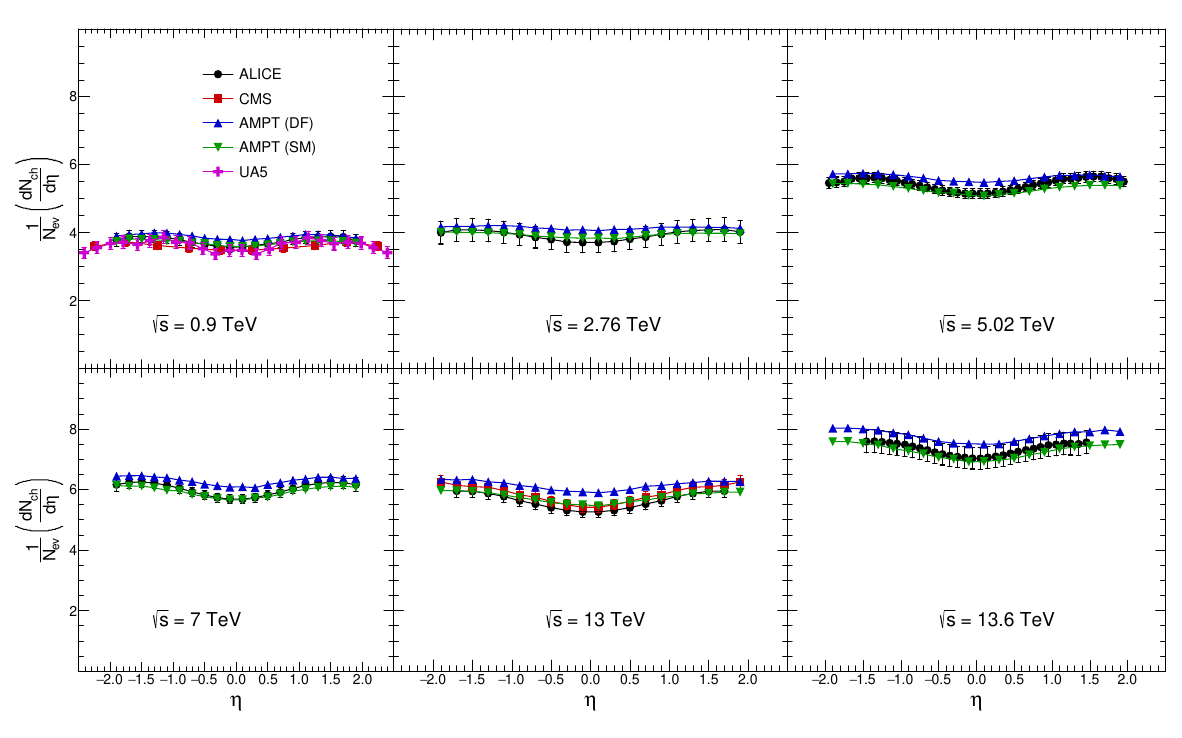}
\caption{Pseudorapidity density of charged particles as a function of $\eta$ in pp collisions at $\sqrt{s}$ = 0.9–13.6 TeV. Results from the AMPT default (DF) and string melting (SM) modes are compared with the corresponding experimental measurements; the references are listed in Table~\ref{tab:exp_refs}. The lines joining the points are drawn to guide the eyes.}
\label{fig1}
\end{center}
\end{figure}

\begin{figure}[!ht]
\begin{center}
\begin{tabular}{cc}
\includegraphics[scale=0.35]{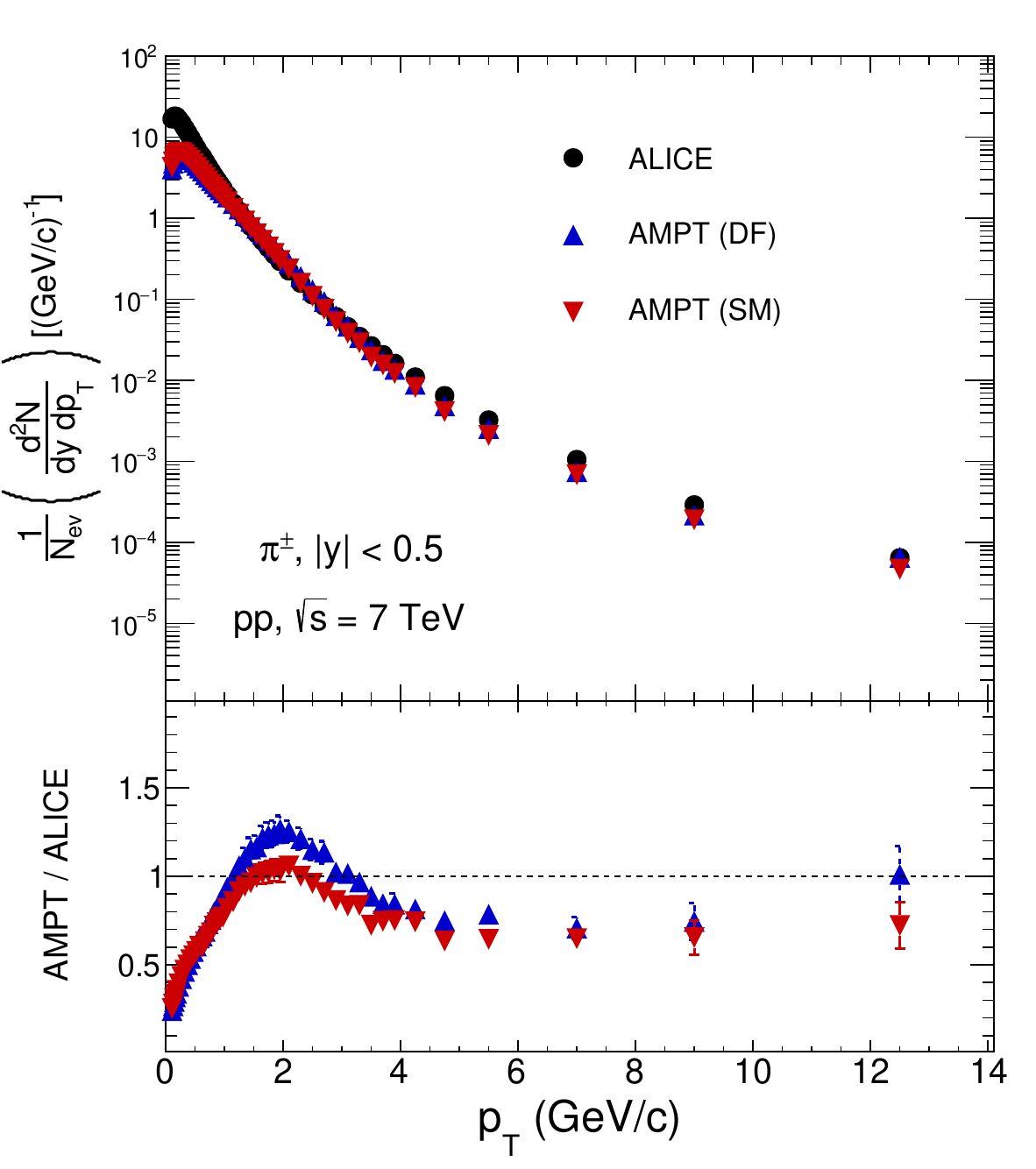} &
\includegraphics[scale=0.35]{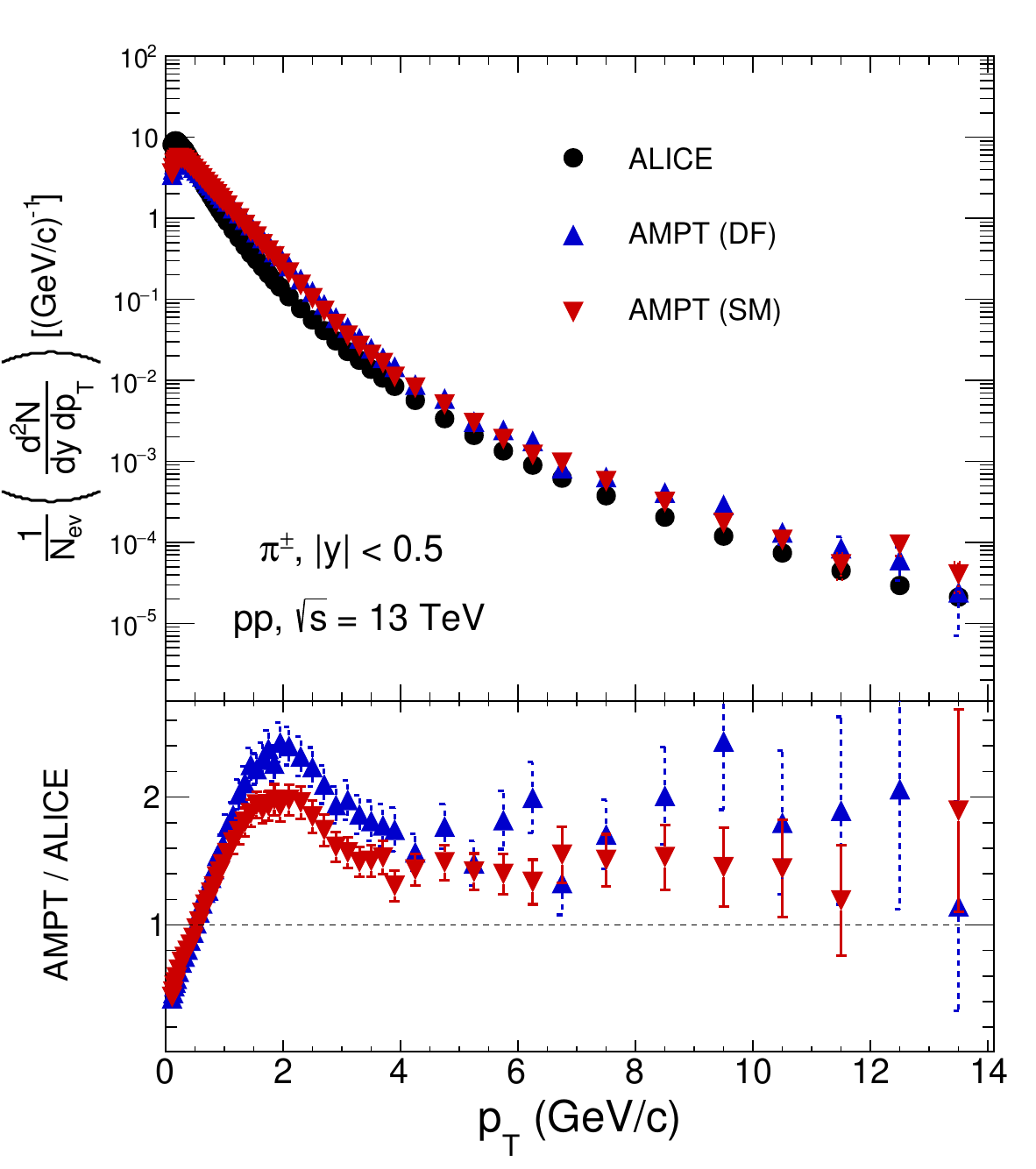} \\
(a) & (b) \\
\end{tabular}
\caption{Transverse momentum distributions of charged pions, $\pi^{\pm}$, produced in $pp$ collisions at (a) $\sqrt{s} = 7$~TeV and (b) $13$~TeV, within the mid-rapidity interval $|y| < 0.5$. Theoretical predictions obtained with the AMPT default (DF) and string melting (SM) modes are presented alongside the corresponding ALICE measurements, whose references are listed in Table~\ref{tab:exp_refs}. The lower panels show the ratio of each model prediction to the measured data.}
\label{fig2}
\end{center}
\end{figure}

\subsubsection{AMPT Default (DF)}
\label{sub_sub_sec:ampt_df}

In the default configuration of AMPT, the initial state is modeled by the HIJING event generator, which produces excited strings from multiple parton--parton scattering. These strings are then fragmented into hadrons using the Lund string model. No partonic scattering phase is included. The resulting hadrons undergo further interactions through a hadronic cascade, which is treated by the ART (A Relativistic Transport) model~\cite{art_ref}. The DF mode therefore provides a description of the collision in which all dynamics proceeds through hadronic degrees of freedom.

\begin{table}[!ht]
\begin{center}
\caption{Event samples generated with AMPT for each centre-of-mass energy.}
\label{tab:event_samples}
\begin{tabular}{cc}
\hline
$\sqrt{s}$ (TeV) & Number of events \\
\hline
$0.9$   & $2.1 \times 10^{6}$ \\
$2.76$  & $2.5 \times 10^{6}$ \\
$5.02$  & $3.9 \times 10^{6}$ \\
$7$     & $6.0 \times 10^{6}$ \\
$13$    & $5.2 \times 10^{6}$ \\
$13.6$  & $6.7 \times 10^{6}$ \\
\hline
\end{tabular}
\end{center}
\end{table}

\subsubsection{AMPT String Melting (SM)}
\label{sub_sub_sec:ampt_sm}

The string melting configuration differs from the default mode in the treatment of the partonic stage. Instead of fragmenting directly into hadrons, the strings produced by HIJING are converted into their valence quarks and antiquarks, which are then propagated through a partonic cascade described by the ZPC (Zhang's Parton Cascade) model~\cite{zpc_ref}. Partons interact through two-body scattering with a cross section controlled by the strong coupling constant and a screening mass. Once partonic interactions cease, quarks and antiquarks are recombined into hadrons via a coalescence mechanism. The resulting hadron gas then evolves through the same hadronic cascade as in the DF mode. Because partonic degrees of freedom and quark coalescence are included, the SM configuration captures dynamical features that are absent in the default mode.

The hadronic cascade time, denoted $\tau_{\rm HC}$, is a parameter of the hadronic stage in AMPT. It is studied exclusively with the SM configuration, for two reasons. First, the hadronic cascade time is meaningful only as a regulator of the hadronic stage, and its effect on the FB correlation strength can be isolated most cleanly when the initial state is generated with the same mechanism in all runs. Varying $\tau_{\rm HC}$ within the DF mode would conflate the effect of the cascade duration with the difference in initial-state description between the two configurations. Second, the partonic phase in the SM mode is expected to be the dominant source of long-range FB correlations, so the hadronic cascade acts as a well-defined dilution stage whose duration can be scanned. This makes the SM mode the natural setting in which to probe how much of the correlation survives hadronic rescattering.

The event samples used in this analysis are summarised in Table~\ref{tab:event_samples}. For each energy, a total of $2.1 \times 10^{6}$ to $6.7 \times 10^{6}$ minimum-bias events were generated with AMPT, split between the DF and SM configurations. The same event sample is used across all analyses at a given energy, except for the cascade-time study, where dedicated runs are performed for each value of $\tau_{\rm HC}$.

\begin{table}[!ht]
\begin{center}
\caption{Experimental data sets used for the validation of the AMPT results. The energy, the observable, and the corresponding reference are listed for each data set.}
\label{tab:exp_refs}
\begin{tabular}{c l l}
\hline
$\sqrt{s}$ (TeV) & Observable & Reference \\
\hline
$0.9$   & Pseudorapidity density  & ALICE~\cite{alice_dndeta_09} \\
	&	& CMS~\cite{cms_dndeta_09}\\
	&	& UA5~\cite{ua5_dndeta_09}\\
	
$2.76$  & Pseudorapidity density & ALICE~\cite{alice_dndeta_09} \\
$5.02$  & Pseudorapidity density & ALICE~\cite{alice_dndeta_502} \\
$7$     & Pseudorapidity density & ALICE~\cite{alice_dndeta_09} \\
$13$    & Pseudorapidity density & ALICE~\cite{alice_dndeta_13} \\
	& & CMS~\cite{cms_dndeta_13} \\
$13.6$  & Pseudorapidity density & ALICE~\cite{alice_dndeta_136} \\
\hline
$7$     & $p_{\rm T}$ spectra of $\pi^{\pm}$ & ALICE~\cite{alice_pt_7} \\
$13$    & $p_{\rm T}$ spectra of $\pi^{\pm}$ & ALICE~\cite{alice_pt_13} \\
\hline
\end{tabular}
\end{center}
\end{table}

To verify that the simulated events reproduce the bulk features of the data, the AMPT results are compared with the available experimental measurements. The charged-particle pseudorapidity density, $\mathrm{d}N_{\rm ch}/\mathrm{d}\eta$, is shown in Figure~\ref{fig1} for all six energies, and the transverse momentum spectra of charged pions at $\sqrt{s} = 7$ and $13$~TeV are shown in Figure~\ref{fig2}. The experimental data sets used in these comparisons are listed in Table~\ref{tab:exp_refs}, together with the corresponding energy and the observable. Agreement within the quoted uncertainties is observed over the kinematic range considered here, which supports the use of the AMPT configurations for the correlation analysis that follows.

\begin{figure}[!ht]
\begin{center}
\begin{tabular}{cc}
\includegraphics[scale=0.15]{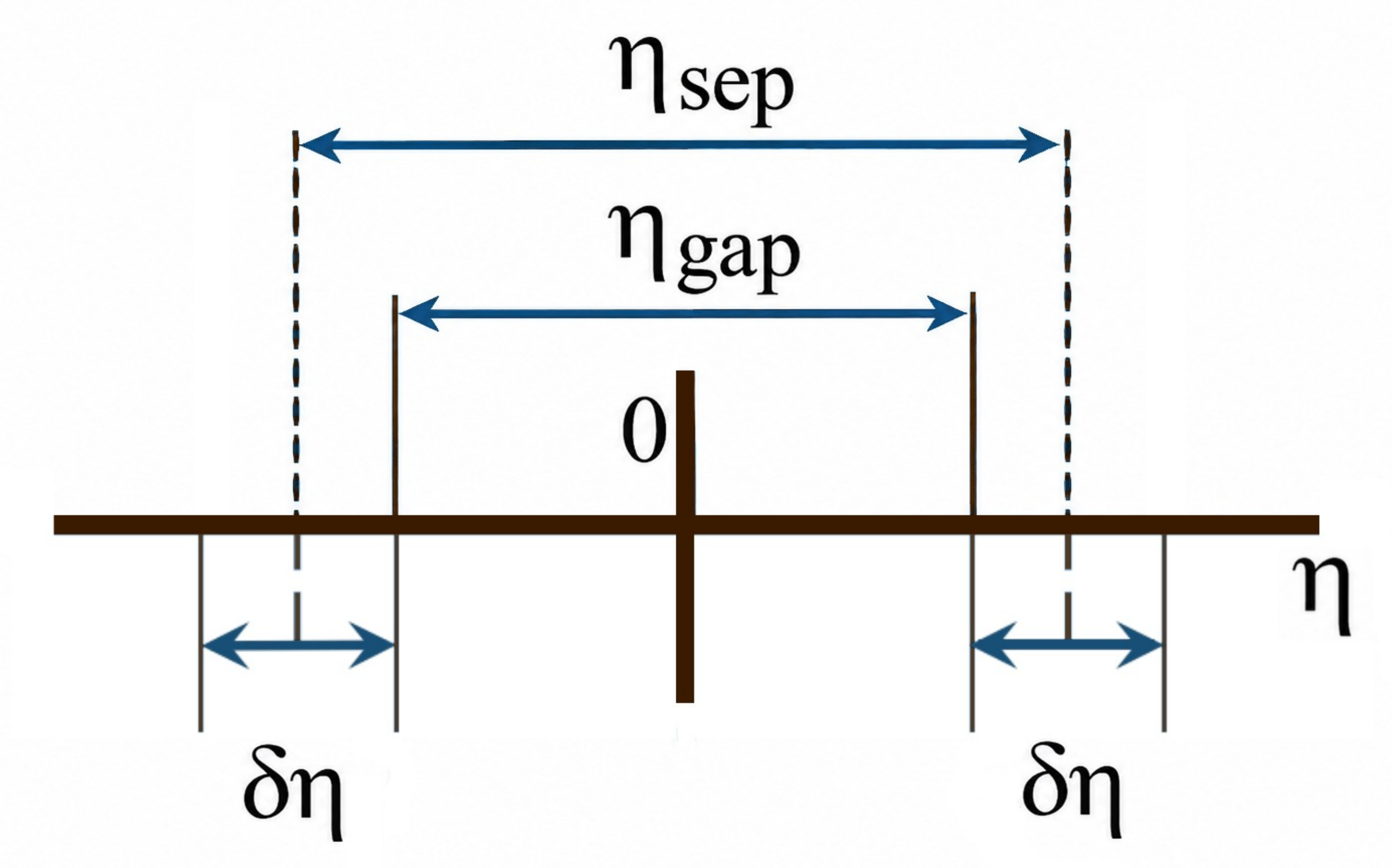} &
\includegraphics[scale=0.30]{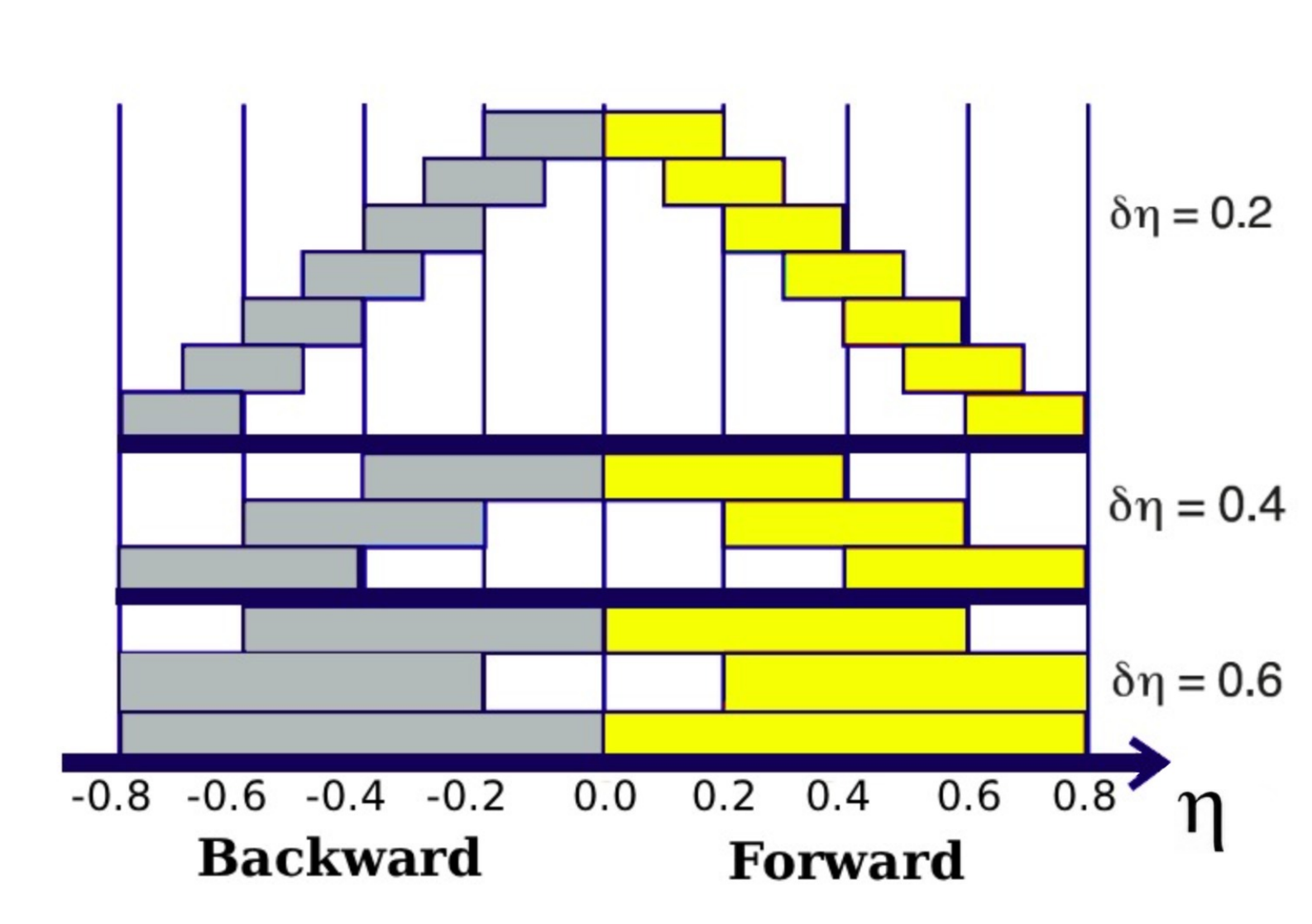} \\
(a) & (b) \\
\end{tabular}
\caption{Schematic representation of the pseudorapidity observables adopted in this study. (a) Definition of the variables $\delta\eta$, $\eta_{\rm gap}$, and $\eta_{\rm sep}$. (b) Arrangement of $\eta$-windows with varying widths $\delta\eta$ and separation gaps $\eta_{\rm gap}$.}
\label{fig3}
\end{center}
\end{figure}

\begin{figure}[!h]
\begin{center}
\includegraphics[scale=0.25]{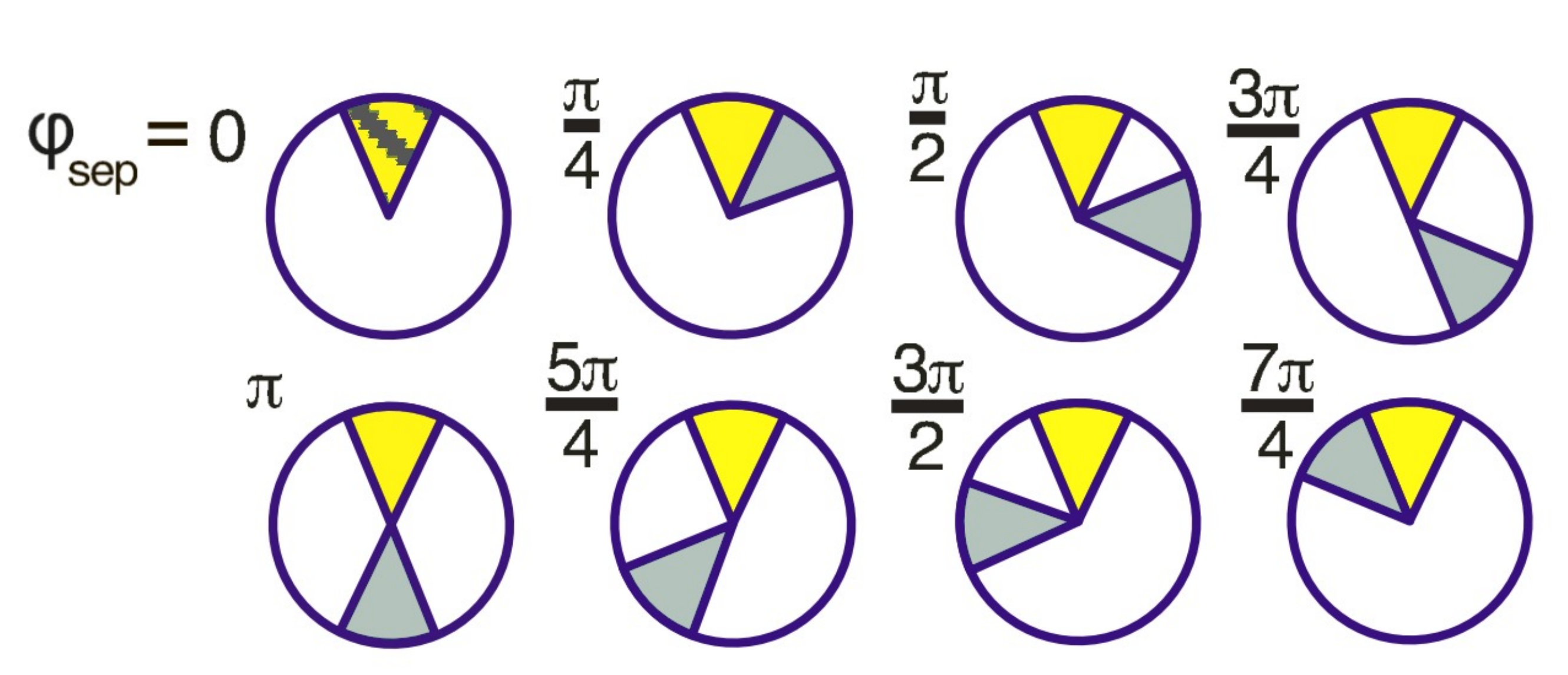}
\caption{Layout of the eight azimuthal sector configurations considered in this analysis. Each forward and backward pseudorapidity window, of width $\delta\eta = 0.2$, is further divided into eight azimuthal sectors of width $\delta\varphi = \pi/4$. The grey sectors belong to the first window of the FB pair and the yellow sectors to the second. The quantity $\varphi_{\rm sep}$ denotes the azimuthal separation between the centres of the two sectors.}
\label{fig4}
\end{center}
\end{figure}

\begin{figure}[!ht]
\begin{center}
\begin{tabular}{cc}
\includegraphics[scale=0.40]{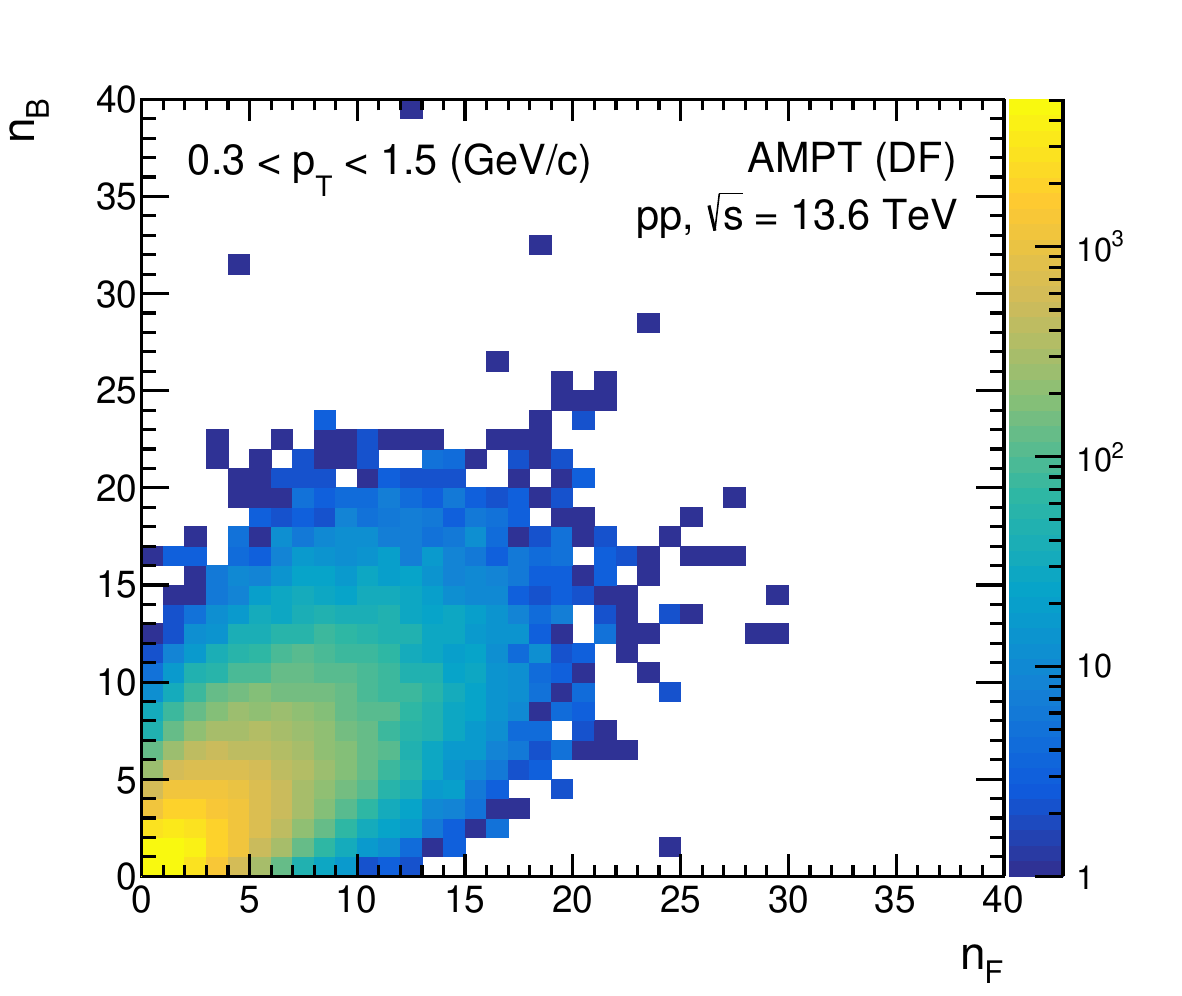} &
\includegraphics[scale=0.40]{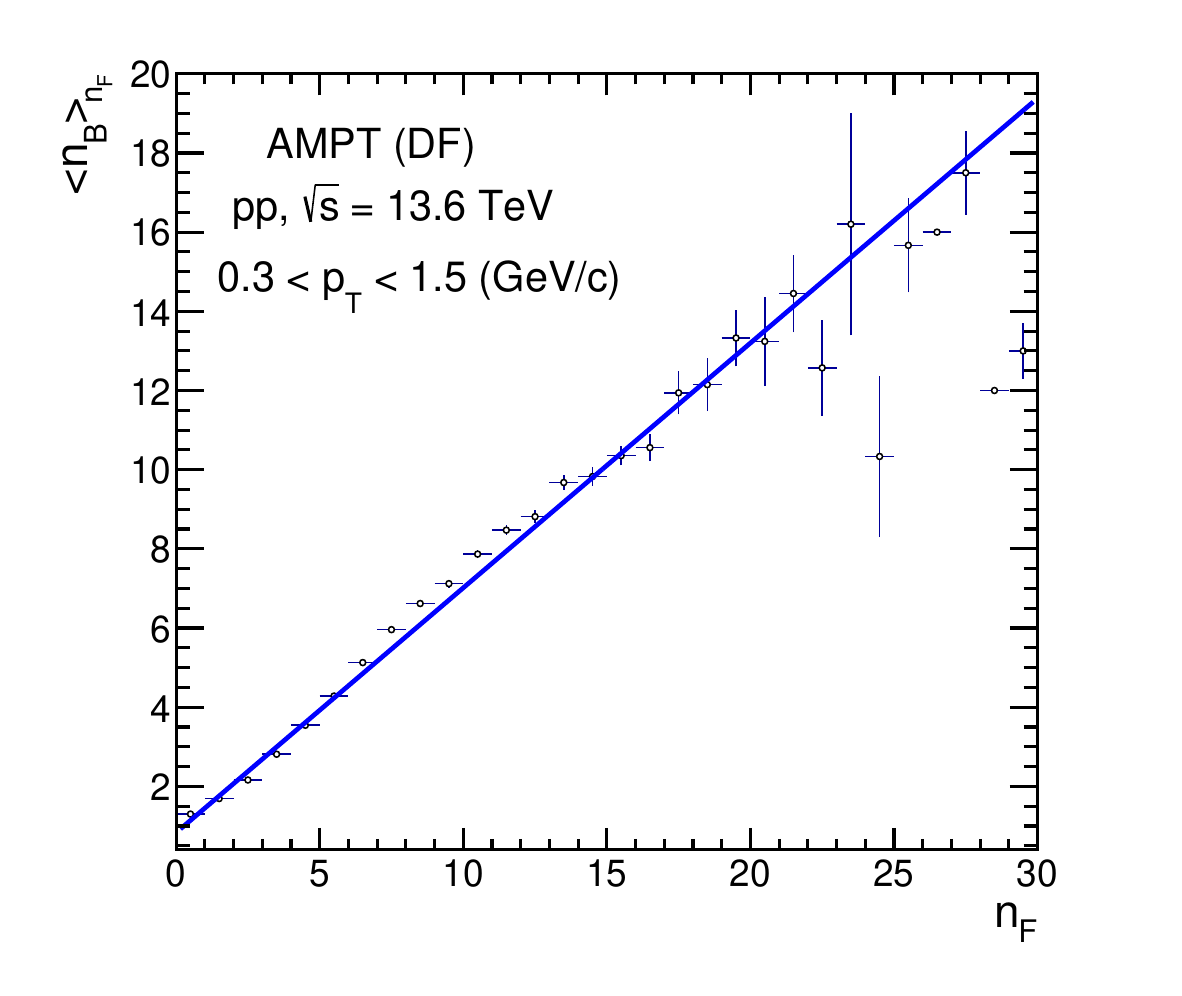} \\
(a) & (b) \\
\includegraphics[scale=0.40]{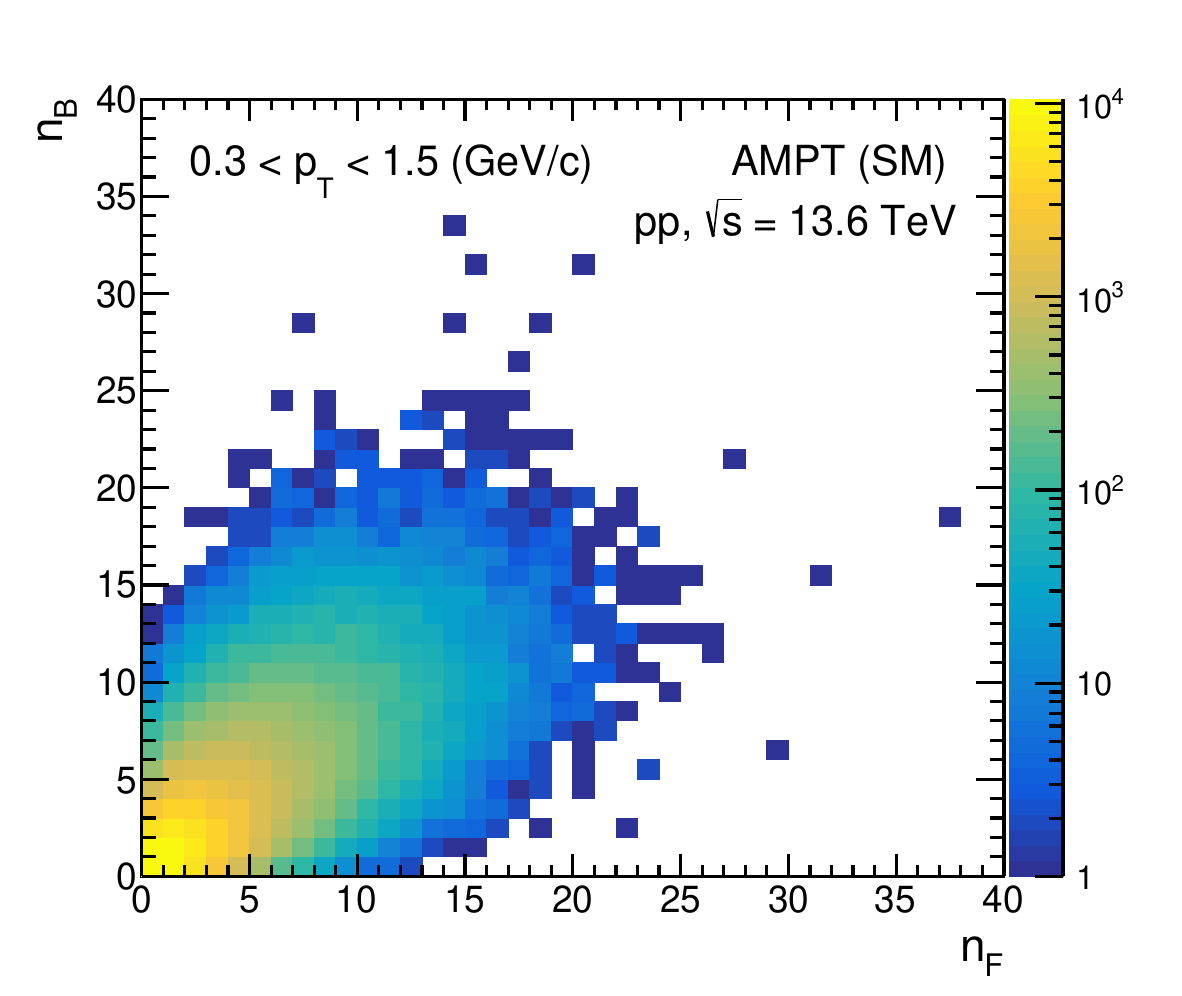} &
\includegraphics[scale=0.40]{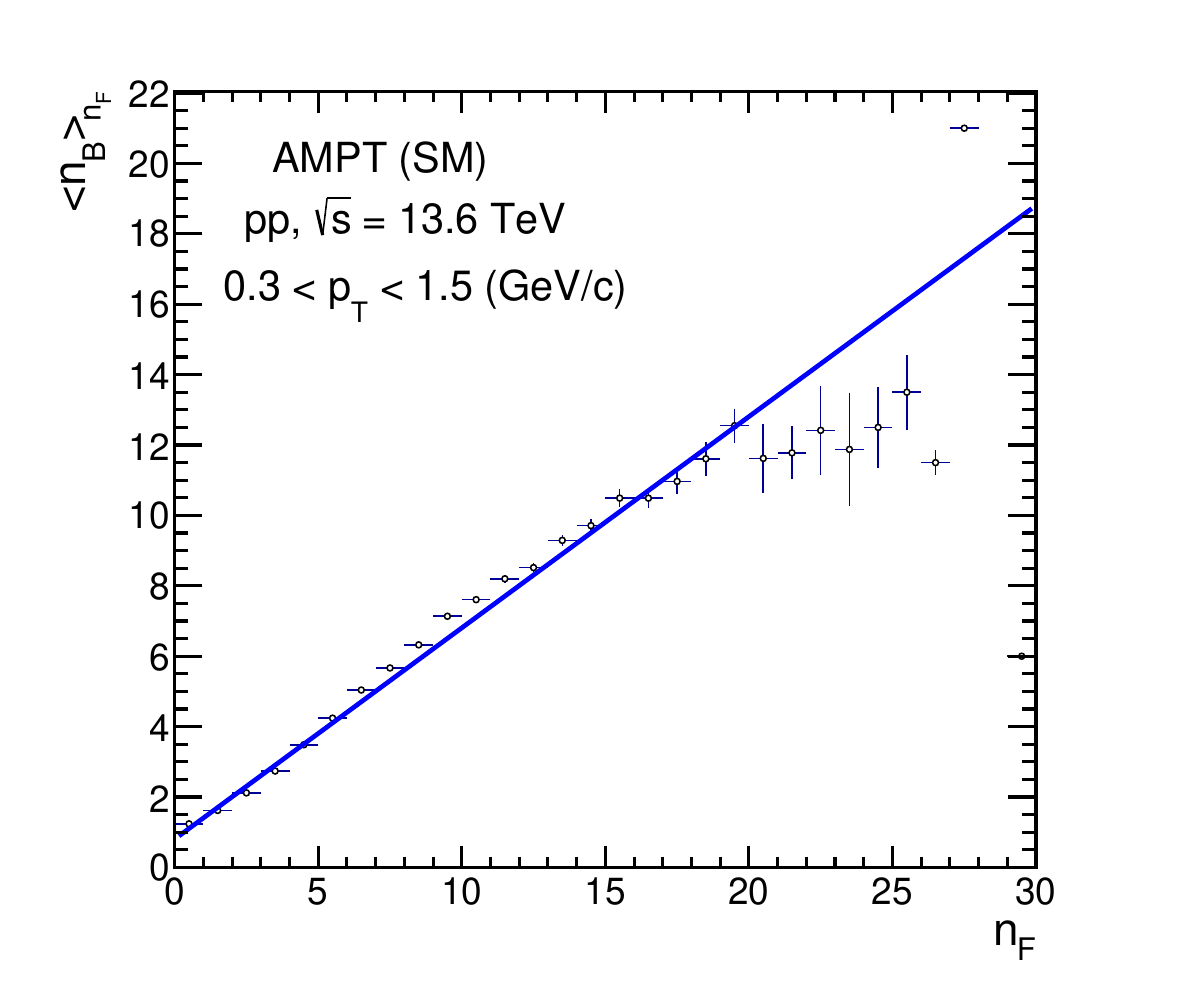} \\
(c) & (d) \\
\end{tabular}
\caption{Representative forward and backward raw multiplicity distributions for windows of width $\delta\eta = 0.8$ and $\eta_{\rm gap} = 0.0$ at $\sqrt{s} = 13.6$ TeV, obtained with the AMPT default mode in (a) and the string melting mode in (c). The corresponding correlation functions are shown in (b) for the default mode and (d) for the string melting mode.}
\label{fig5}
\end{center}
\end{figure}

\subsection{Definition of counting windows}
\label{sub_sec:count_windows}

Two intervals of variable width $\delta\eta$, ranging from $0.2$ to $0.8$ and placed symmetrically about $\eta = 0$, are labelled as forward ($\mathrm{F}$, $\eta > 0$) and backward ($\mathrm{B}$, $\eta < 0$). Correlations between the multiplicities of charged particles, $n$, are then examined as a function of the gap between the two windows, denoted $\eta_{\rm gap}$. A second and more convenient variable is $\eta_{\rm sep}$, the pseudorapidity separation between the centres of the windows. Both variables are illustrated in Figure~\ref{fig3}(a), while all configurations of window pairs adopted in the analysis are shown in Figure~\ref{fig3}(b). The study is extended to correlations between separated regions in the $\eta$--$\varphi$ plane, referred to as sectors. The azimuthal angle is divided into eight sectors of width $\delta\varphi = \pi/4$, as shown in Figure~\ref{fig4}. This choice reflects a compromise between granularity and statistical uncertainty. The definitions and relations given in Section~\ref{sec:intro} carry over unchanged to the $\eta$--$\varphi$ windows. 

\subsection{Procedures of the FB correlation coefficient measurement}

The present work is devoted to FB correlations associated with soft particle production. Accordingly, the transverse momentum is restricted to $0.3 < p_{\rm T} < 1.5$~GeV/$c$, with the single exception of the $p_{\rm T}$-dependence study in Section~\ref{sec:pt}, where the range is widened to $0.3 < p_{\rm T} < 6$~GeV/$c$.

Two procedures are available for extracting the correlation coefficient $b_{\rm corr}$ for a given window pair~\cite{ref1,ref2,ref3,ref4}. In the first, the event-by-event averages $\langle n_{\rm B} n_{\rm F} \rangle$, $\langle n_{\rm B} \rangle$, $\langle n_{\rm F} \rangle$, and $\langle n_{\rm F}^{2} \rangle$ are accumulated and $b_{\rm corr}$ is then evaluated from Eq.~\eqref{eq:bcorr_pearson}. In the second, $b_{\rm corr}$ is obtained by linear regression: the two-dimensional distribution of $(n_{\rm B}, n_{\rm F})$ is constructed from all selected events, the mean backward multiplicity is determined for each fixed value of the forward multiplicity, and $b_{\rm corr}$ follows from a linear fit to the resulting correlation function, as illustrated in Figure~\ref{fig5}. Departures from linearity could in principle carry additional information, but a detailed investigation of such effects lies outside the scope of this article.

The two procedures have been shown to yield results that agree within statistical uncertainties. The results presented here are obtained with the first method.

\begin{figure}[!h]
\begin{center}
\includegraphics[scale=0.75]{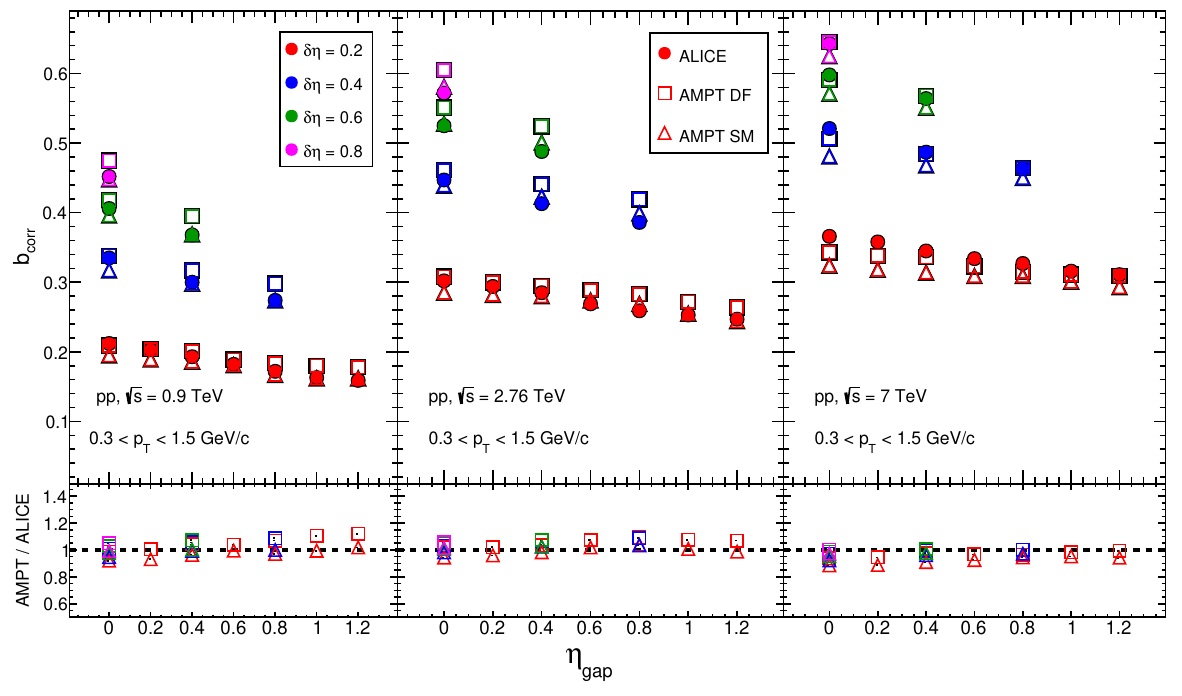}
\caption{Forward-backward correlation strength $b_{\rm corr}$ as a function of the pseudorapidity gap $\eta_{\rm gap}$ for window widths $\delta\eta = 0.2$, $0.4$, $0.6$, and $0.8$ in pp collisions at $\sqrt{s} = 0.9$, $2.76$, and $7$ TeV. Results from the AMPT default (DF) and string melting (SM) modes are compared with the ALICE measurements. Filled symbols denote the ALICE data \cite{alice_main} and open symbols the AMPT results, with $\delta\eta = 0.2$ (red), $0.4$ (blue), $0.6$ (green), and $0.8$ (magenta). The lower panels display the corresponding ratio plots.}
\label{fig6}
\end{center}
\end{figure}

\begin{figure}[!h]
\begin{center}
\includegraphics[scale=0.75]{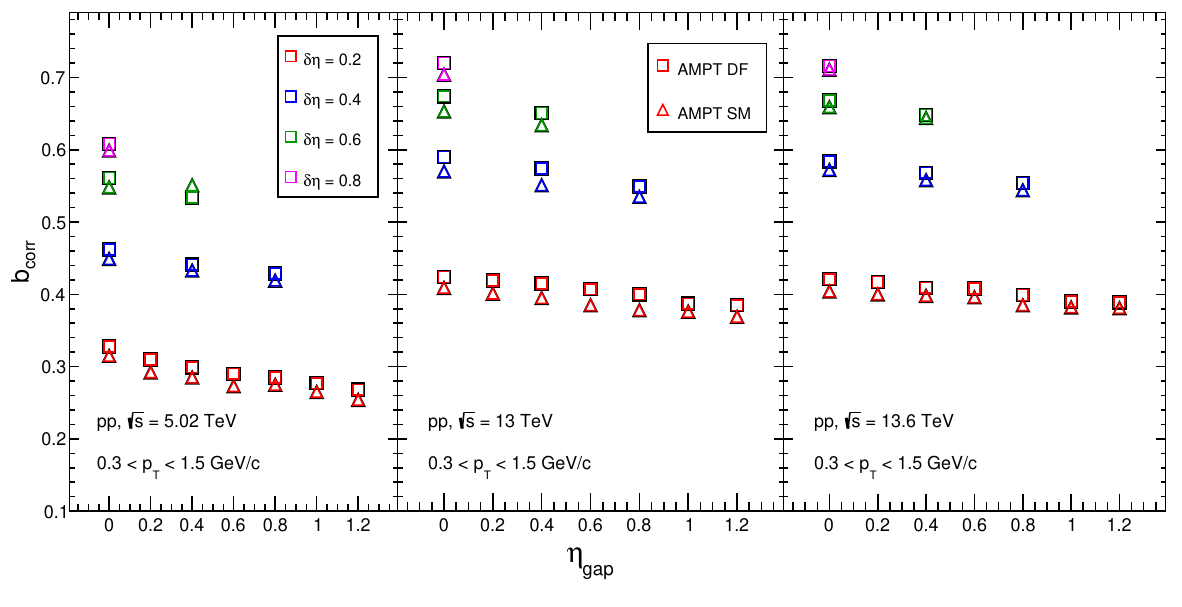}
\caption{Forward-backward correlation strength $b_{\rm corr}$ as a function of the pseudorapidity gap $\eta_{\rm gap}$ for window widths $\delta\eta = 0.2$, $0.4$, $0.6$, and $0.8$ in pp collisions at $\sqrt{s} = 5.02$, $13$, and $13.6$ TeV. Results from the AMPT default (DF) and string melting (SM) modes are shown, with $\delta\eta = 0.2$ (red), $0.4$ (blue), $0.6$ (green), and $0.8$ (magenta).}
\label{fig7}
\end{center}
\end{figure}

\section{Multiplicity correlations in pseudorapidity-separated windows}
\label{sec:mult_sep}

\subsection{Dependence on the gap between windows}
\label{sub_sec:window_gap}

Figures~\ref{fig6} and~\ref{fig7} show the FB multiplicity correlation coefficient $b_{\rm corr}$ as a function of $\eta_{\rm gap}$ for different window widths $\delta\eta$ in $pp$ collisions at $\sqrt{s} = 0.9$--$13.6$ TeV. At each energy, $b_{\rm corr}$ is found to decrease gradually with increasing $\eta_{\rm gap}$, while retaining a substantial pedestal value across the full $\eta_{\rm gap}$ range. In Figure~\ref{fig6}, the results for $\sqrt{s} = 0.9$, $2.76$, and $7$ TeV are compared with the corresponding ALICE measurements \cite{alice_main}, and the lower panels display the ratio of the model predictions to the data. Across all energies considered, the AMPT default (DF) mode is seen to produce a stronger correlation than the string melting (SM) mode. At lower energies, namely $0.9$ and $2.76$ TeV, the SM mode reproduces the ALICE measurements \cite{alice_main} better, whereas at $7$ TeV the DF mode provides the closer description. This energy-dependent switch in the relative performance of the two configurations suggests that both the partonic and hadronic stages of the collision contribute to the observed correlations, and that neither alone is sufficient to account for the full energy dependence. A plausible origin of this behaviour is that the relative weight of partonic scattering and hadronic rescattering changes with collision energy, so that the configuration that best captures the dominant mechanism at a given energy also changes.

\begin{figure}[!h]
\begin{center}
\includegraphics[scale=0.75]{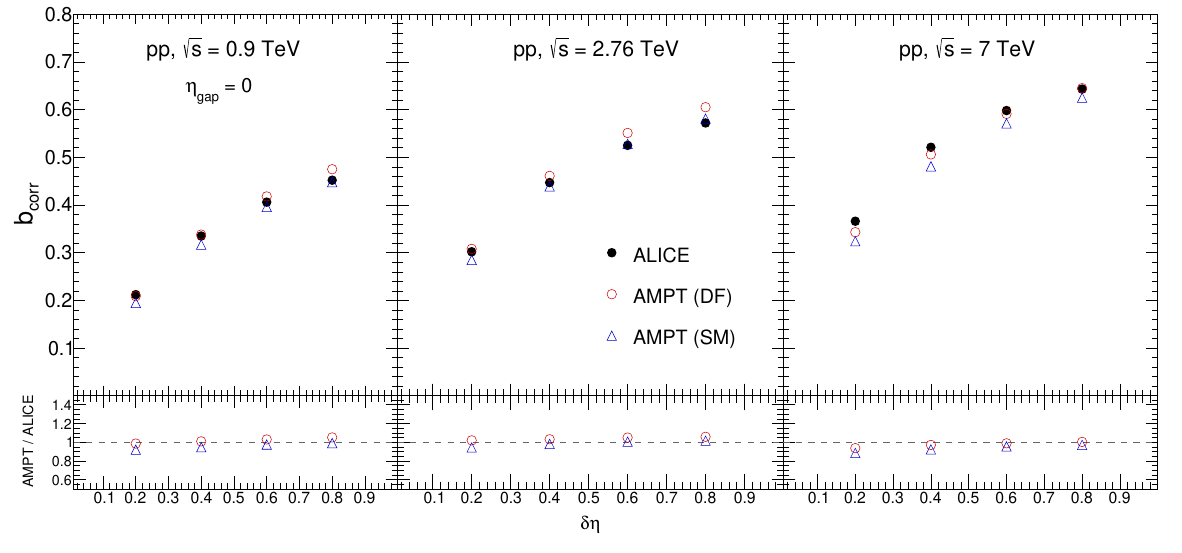}
\caption{Correlation strength $b_{\rm corr}$ as a function of the window width $\delta\eta$ for a pseudorapidity gap $\eta_{\rm gap} = 0$ in pp collisions at $\sqrt{s} = 0.9$, $2.76$, and $7$ TeV. Results from the AMPT string melting (open triangles) and default (open circles) modes are compared with the ALICE measurements \cite{alice_main}. The lower panels display the ratio of $b_{\rm corr}$ between ALICE and AMPT.}
\label{fig8}
\end{center}
\end{figure}

\subsection{Dependence on the width of windows}
\label{sub_sec:window_width}

The dependence on $\delta\eta$ for adjacent windows ($\eta_{\rm gap} = 0$), placed symmetrically about $\eta = 0$, is displayed in Figure~\ref{fig8}. At $\sqrt{s} = 0.9$, $2.76$, and $7$ TeV, the correlation coefficient rises non-linearly with $\delta\eta$ in both the AMPT default (DF) and string melting (SM) configurations. The results are compared with the ALICE measurements \cite{alice_main}, and the lower panels show the corresponding ratio plots. This behaviour is reproduced reasonably well by both configurations at all energies considered.

\subsection{Dependence on the collision energy}
\label{sub_sec:collision_energy}

Figures~\ref{fig6} and~\ref{fig7} show that the pedestal value of $b_{\rm corr}$ grows with $\sqrt{s}$, whereas the slope of the $b_{\rm corr}(\eta_{\rm gap})$ dependence remains approximately constant. This behaviour is observed in both the AMPT (DF) and AMPT (SM) configurations. It suggests that the short-range contribution carries only a weak dependence on $\sqrt{s}$, while the long-range multiplicity correlations dominate in $pp$ collisions and their strength rises appreciably with $\sqrt{s}$. This rise cannot be accounted for by the growth of the mean multiplicity alone: when window sizes are chosen so that the mean multiplicity is held fixed across energies, the increase persists. Our findings are consistent with the ALICE measurements~\cite{alice_main}.

A pronounced energy dependence and rather large values of $b_{\rm corr}$ were reported earlier by the UA5 collaboration~\cite{ref3} and more recently by the ATLAS collaboration~\cite{ref16}. As Figure~\ref{fig6} demonstrates, however, the correlation coefficient varies non-linearly with the width of the pseudorapidity window. This must be taken into account when comparing correlation strengths extracted under different experimental conditions. In particular, it accounts for the small values of $b_{\rm corr}$ measured by the STAR collaboration at RHIC in $pp$ collisions at $\sqrt{s} = 200$ GeV~\cite{ref17}, where narrow FB windows ($\delta\eta = 0.2$) were used, whereas earlier $pp$ and $p\bar{p}$ experiments employed windows of a few units of pseudorapidity.

\begin{figure}[!h]
\begin{center}
\includegraphics[scale=0.75]{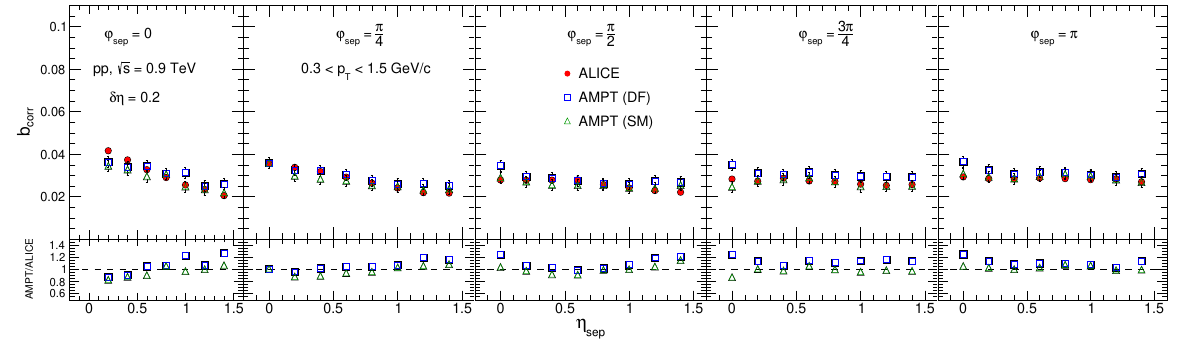}
\caption{Correlation strength $b_{\rm corr}$ for separated $\eta$--$\varphi$ window pairs in pp collisions at $\sqrt{s} = 0.9$ TeV as a function of the pseudorapidity separation, for fixed window widths $\delta\eta = 0.2$ and $\delta\varphi = \pi/4$. The panels correspond to different azimuthal separation distances between the two sectors: $\varphi_{\rm sep} = 0$, $\pi/4$, $\pi/2$, $3\pi/4$, and $\pi$. Results from the AMPT default (open squares) and string melting (open triangles) modes are compared with the ALICE measurements \cite{alice_main}(filled circles). The lower panels display the ratio of $b_{\rm corr}$ between the ALICE and AMPT results.}
\label{fig9}
\end{center}
\end{figure}

\begin{figure}[!h]
\begin{center}
\includegraphics[scale=0.75]{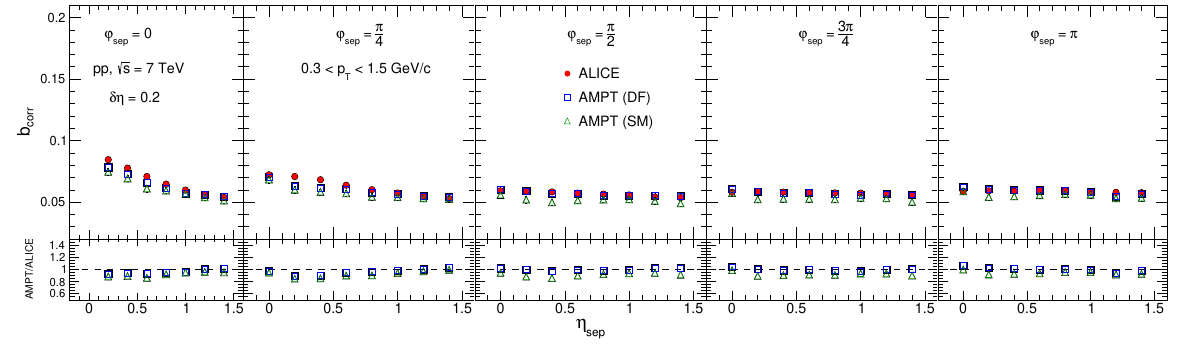}
\caption{Correlation strength $b_{\rm corr}$ for separated $\eta$--$\varphi$ window pairs in pp collisions at $\sqrt{s} = 7$ TeV as a function of the pseudorapidity separation, for fixed window widths $\delta\eta = 0.2$ and $\delta\varphi = \pi/4$. Results from the AMPT default (open squares) and string melting (open triangles) modes are compared with the ALICE measurements \cite{alice_main}(filled circles). The lower panels display the ratio of $b_{\rm corr}$ between the ALICE and AMPT results.}
\label{fig10}
\end{center}
\end{figure}

\begin{figure}[!h]
\begin{center}
\includegraphics[scale=0.75]{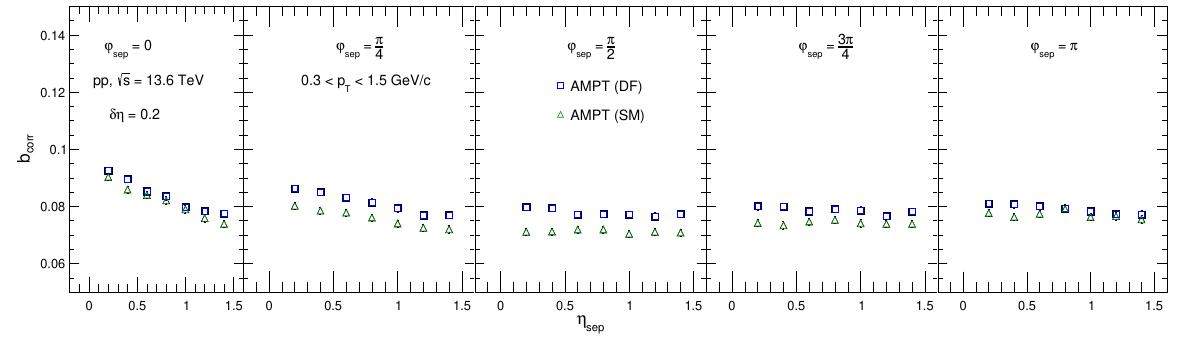}
\caption{Correlation strength $b_{\rm corr}$ for separated $\eta$--$\varphi$ window pairs in pp collisions at $\sqrt{s} = 13.6$ TeV as a function of the pseudorapidity separation, for fixed window widths $\delta\eta = 0.2$ and $\delta\varphi = \pi/4$. Results from the AMPT default (open squares) are presented. }
\label{fig11}
\end{center}
\end{figure}

\section{Multiplicity correlations in separated $\eta$ and $\varphi$ windows}
\label{sec:eta_phi}

Multiplicity correlations are also examined for different configurations of forward and backward azimuthal sectors. These sectors are defined within separated forward and backward pseudorapidity windows of width $\delta\eta = 0.2$, each divided into azimuthal sectors of width $\delta\varphi = \pi/4$, as shown in Figure~\ref{fig4}. This yields five sector pairs corresponding to different values of the azimuthal separation. Figures~\ref{fig9},~\ref{fig10}, and~\ref{fig11} show the azimuthal dependence of $b_{\rm corr}$ as a function of $\eta_{\rm sep}$ at $\sqrt{s} = 0.9$, $7$, and $13.6$ TeV, respectively.

In all cases considered, the AMPT default (DF) configuration yields higher values of $b_{\rm corr}$ than the string melting (SM) configuration. In Figures~\ref{fig9} and~\ref{fig10}, the results are compared with the ALICE measurements \cite{alice_main}. At $\sqrt{s} = 0.9$ TeV, the SM configuration matches the ALICE data \cite{alice_main} more closely, whereas at $\sqrt{s} = 7$ TeV the DF configuration provides the better description for almost all sector pairs. The corresponding ratio plots are shown in the lower panels of these figures.

The relative performance of the two configurations can be understood in terms of the short- and long-range contributions to $b_{\rm corr}$. The DF mode, which lacks a partonic scattering stage, generates correlations primarily through string fragmentation and hadronic rescattering, a combination that favours shorter-range contributions. The SM mode, by contrast, includes partonic scattering and quark coalescence, which are expected to enhance the long-range component. Since the long-range component is diluted by the hadronic cascade, the net $b_{\rm corr}$ in SM is systematically lower than in DF at the same energy. The change in which configuration agrees better with the data between $0.9$ and $7$ TeV may therefore reflect a shift in the balance between short- and long-range contributions as the collision energy increases.

\begin{figure}[!ht]
\begin{center}
\begin{tabular}{cc}
\includegraphics[scale=0.45]{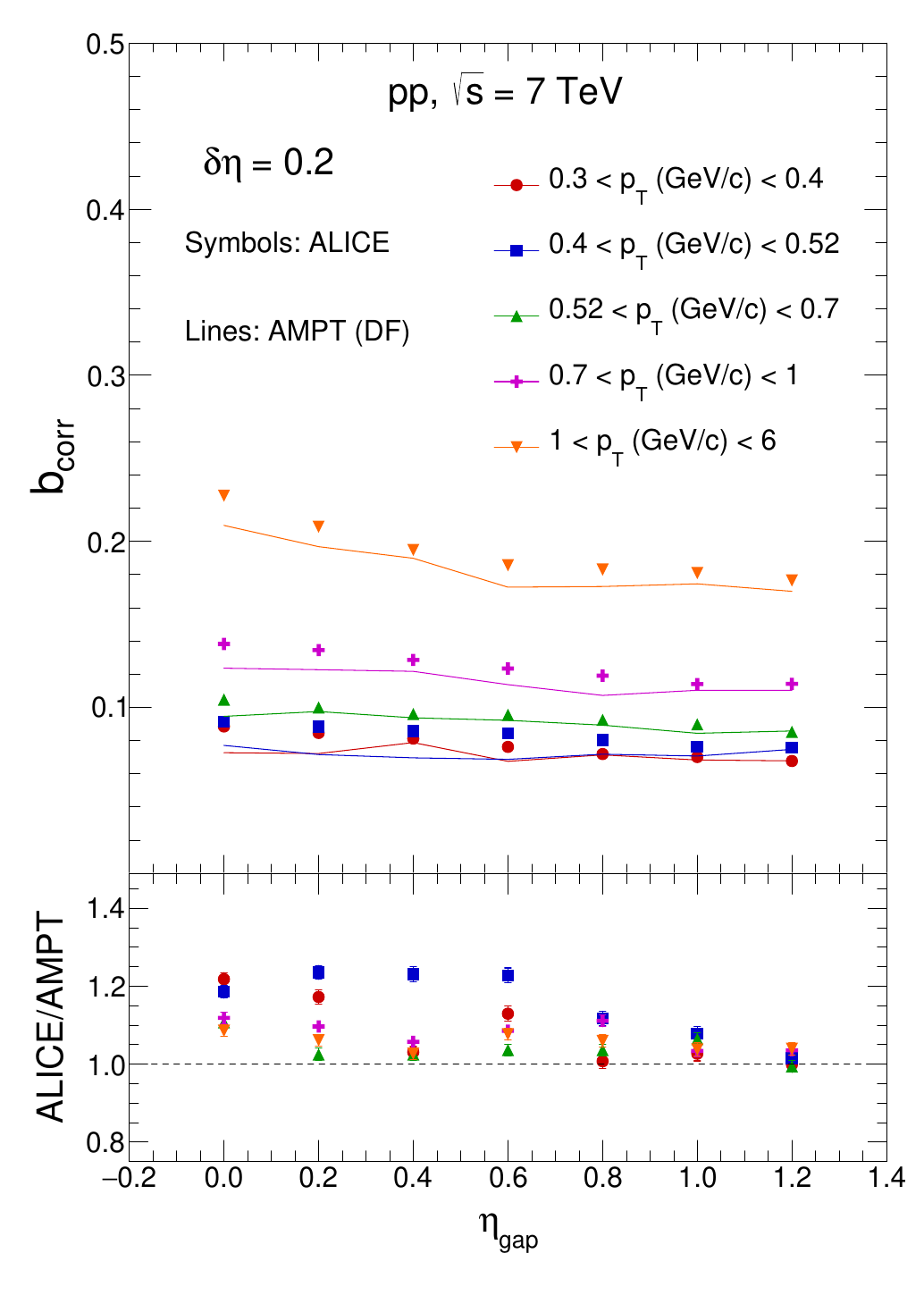} &
\includegraphics[scale=0.45]{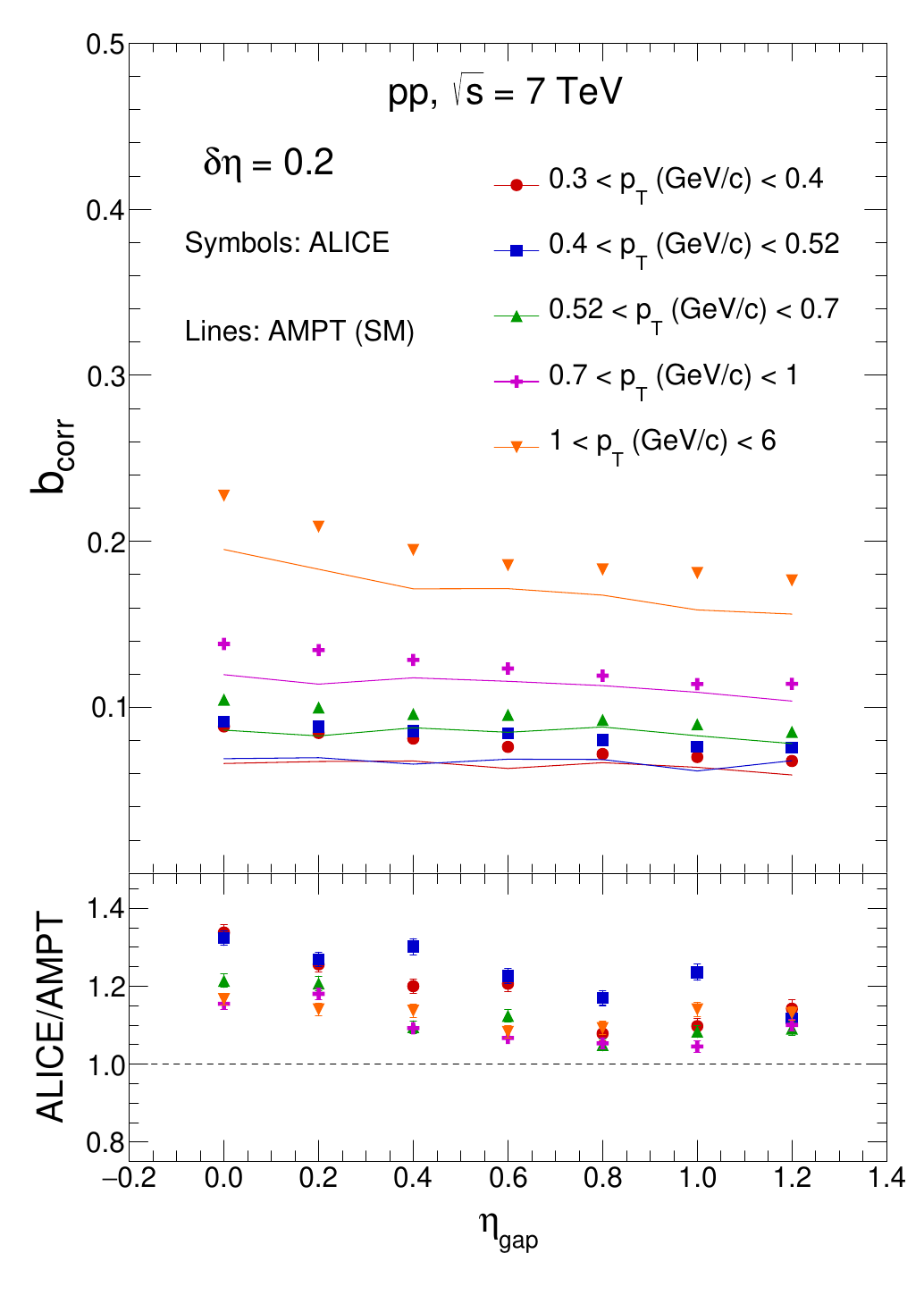} \\
(a) & (b) \\
\end{tabular}
\caption{Correlation strength $b_{\rm corr}$ in pp collisions at $\sqrt{s} = 7$ TeV for separated pseudorapidity window pairs, measured in $p_{\rm T}$ intervals with the same $\langle n_{\rm ch} \rangle$, as a function of the pseudorapidity gap $\eta_{\rm gap}$. The window width is fixed at $\delta\eta = 0.2$. The left and right panels contain the same data points of ALICE measurements \cite{alice_main}, with the lines corresponding to the AMPT default (left) and string melting (right) modes.}
\label{fig12}
\end{center}
\end{figure}

\begin{figure}[!h]
\begin{center}
\includegraphics[scale=0.85]{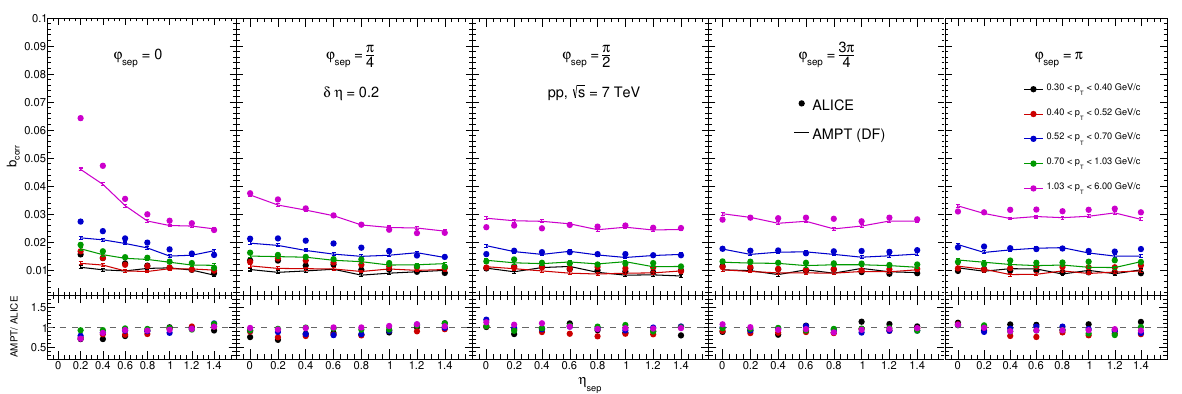}
\caption{Correlation strength $b_{\rm corr}$ in pp collisions at $\sqrt{s} = 7$ TeV for separated $\eta$--$\phi$ windows in different $p_{\rm T}$ intervals with the same $\langle n_{\rm ch} \rangle$. Five values of $\phi_{\rm sep}$ are shown as a function of the pseudorapidity separation $\eta_{\rm sep}$, with window widths $\delta\eta = 0.2$ and $\delta\phi = \pi/4$. The upper panels contain the same ALICE experimental data \cite{alice_main} as symbols, while the lines correspond to the AMPT default (DF) results. The lower panels display the corresponding ratio plots.}
\label{fig13}
\end{center}
\end{figure}

\begin{figure}[!h]
\begin{center}
\includegraphics[scale=0.85]{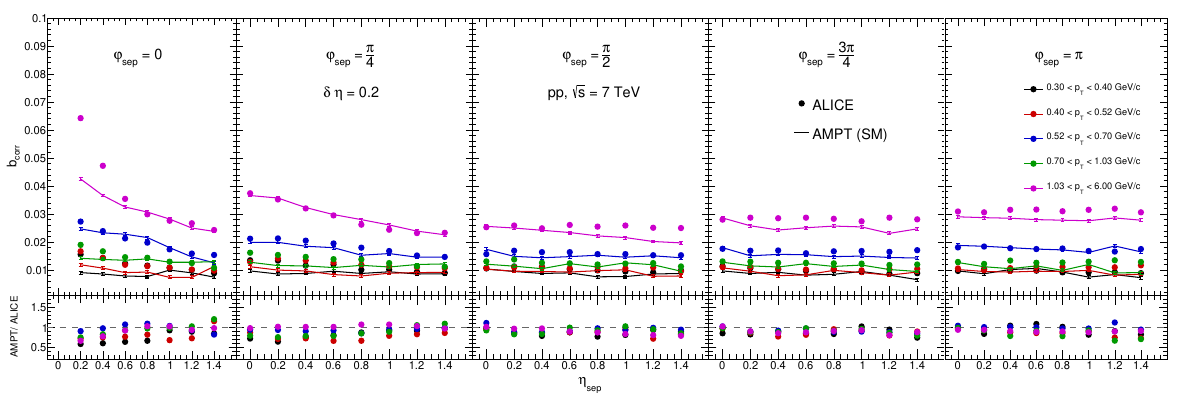}
\caption{Correlation strength $b_{\rm corr}$ in pp collisions at $\sqrt{s} = 7$ TeV for separated $\eta$--$\phi$ windows in different $p_{\rm T}$ intervals with the same $\langle n_{\rm ch} \rangle$. Five values of $\phi_{\rm sep}$ are shown as a function of the pseudorapidity separation $\eta_{\rm sep}$, with window widths $\delta\eta = 0.2$ and $\delta\phi = \pi/4$. The upper panels contain the same ALICE experimental data \cite{alice_main} as symbols, while the lines correspond to the AMPT string melting(SM) results. The lower panels display the corresponding ratio plots.}
\label{fig14}
\end{center}
\end{figure}

\section{FB multiplicity correlations in different $p_T$ intervals}
\label{sec:pt}

The behaviour of the FB multiplicity correlation strength was also examined as a function of the transverse momentum of the registered particles. The correlation coefficient $b_{\rm corr}$ was studied in five $p_{\rm T}$ intervals within $0.3 < p_{\rm T} < 6$~GeV/$c$ at $\sqrt{s} = 7$~TeV: $0.3$--$0.4$, $0.4$--$0.52$, $0.52$--$0.7$, $0.7$--$1.03$, and $1.03$--$6.0$~GeV/$c$, for both the AMPT default (DF) and string melting (SM) configurations. Figure~\ref{fig12} shows $b_{\rm corr}$ as a function of $\eta_{\rm gap}$ for the different $p_{\rm T}$ intervals. In Figure~\ref{fig12}(a), the AMPT DF results are compared with the ALICE measurements \cite{alice_main}, while Figure~\ref{fig12}(b) shows the corresponding comparison for AMPT SM. The ratio plots are displayed in the lower panels of each figure.

In general, the AMPT DF configuration describes the ALICE data \cite{alice_main} better than the SM configuration. The largest deviations for both configurations appear in the lowest $p_{\rm T}$ intervals, around $0.3$--$0.4$~GeV/$c$, where the model predictions are furthest from the measurements.

This behaviour can be understood in terms of the transverse-momentum dependence of the short-range (SR) and long-range (LR) contributions to $b_{\rm corr}$. The ALICE analysis found that both the SR and LR components increase with $p_{\rm T}$, with the LR contribution exhibiting a considerably stronger $p_{\rm T}$ dependence than the SR component~\cite{alice_main}. In the AMPT framework, the SR component is associated with fragmentation of individual sources, such as cluster and resonance decays and jet-like correlations, while the LR component arises from fluctuations in the number and properties of the emitting sources. The low-$p_{\rm T}$ region is dominated by soft, non-perturbative production, where the SR contribution is expected to be relatively more important. The fact that the largest deviations occur at low $p_{\rm T}$ suggests that the description of the SR component in AMPT, particularly its dependence on the emission of soft particles, is less complete than the description of the LR component. At higher $p_{\rm T}$, where the LR contribution dominates and is better constrained by the model's treatment of partonic scattering and string dynamics, the agreement with the data improves.

The combined dependence on transverse momentum and azimuthal separation is examined in Figures~\ref{fig13} and~\ref{fig14}, which show the AMPT default (DF) and string melting (SM) results, respectively, at $\sqrt{s} = 7$~TeV for the same five $p_{\rm T}$ intervals as in Figure~\ref{fig12}. In each figure, five values of $\varphi_{\rm sep}$ are displayed as a function of $\eta_{\rm sep}$, with $\delta\eta = 0.2$ and $\delta\varphi = \pi/4$. The upper panels contain the ALICE measurements \cite{alice_main} as symbols and the AMPT results as lines, and the lower panels show the ratio plots.

For both configurations, $b_{\rm corr}$ decreases with increasing $\eta_{\rm sep}$ at every $\varphi_{\rm sep}$. The dependence on $\varphi_{\rm sep}$ is weaker than on $\eta_{\rm sep}$, the curves for different $\varphi_{\rm sep}$ lie close together, whereas the variation with $\eta_{\rm sep}$ is more pronounced. This indicates that the correlation strength is governed primarily by the longitudinal separation between the windows, with the azimuthal separation playing a secondary role.

The $p_{\rm T}$ ordering seen in Figure~\ref{fig12} persists in the $\eta$--$\varphi$ case. The largest values of $b_{\rm corr}$ occur in the highest $p_{\rm T}$ intervals and the smallest in the lowest, for every $\varphi_{\rm sep}$ and every $\eta_{\rm sep}$, confirming that the $p_{\rm T}$ dependence is not an artefact of the azimuthal geometry.

The DF results lie closer to the ALICE data \cite{alice_main} than the SM results, in agreement with Figure~\ref{fig12}. The deviations are largest at low $p_{\rm T}$ and small $\varphi_{\rm sep}$, and decrease as $\varphi_{\rm sep}$ approaches $\pi$. This is consistent with the interpretation above, the short-range component, dominant at low $p_{\rm T}$ and small azimuthal separation, is less completely described in AMPT, whereas the long-range component, which survives at large separations, is better constrained.

\begin{figure}[!ht]
\begin{center}
\begin{tabular}{cc}
\includegraphics[scale=0.4]{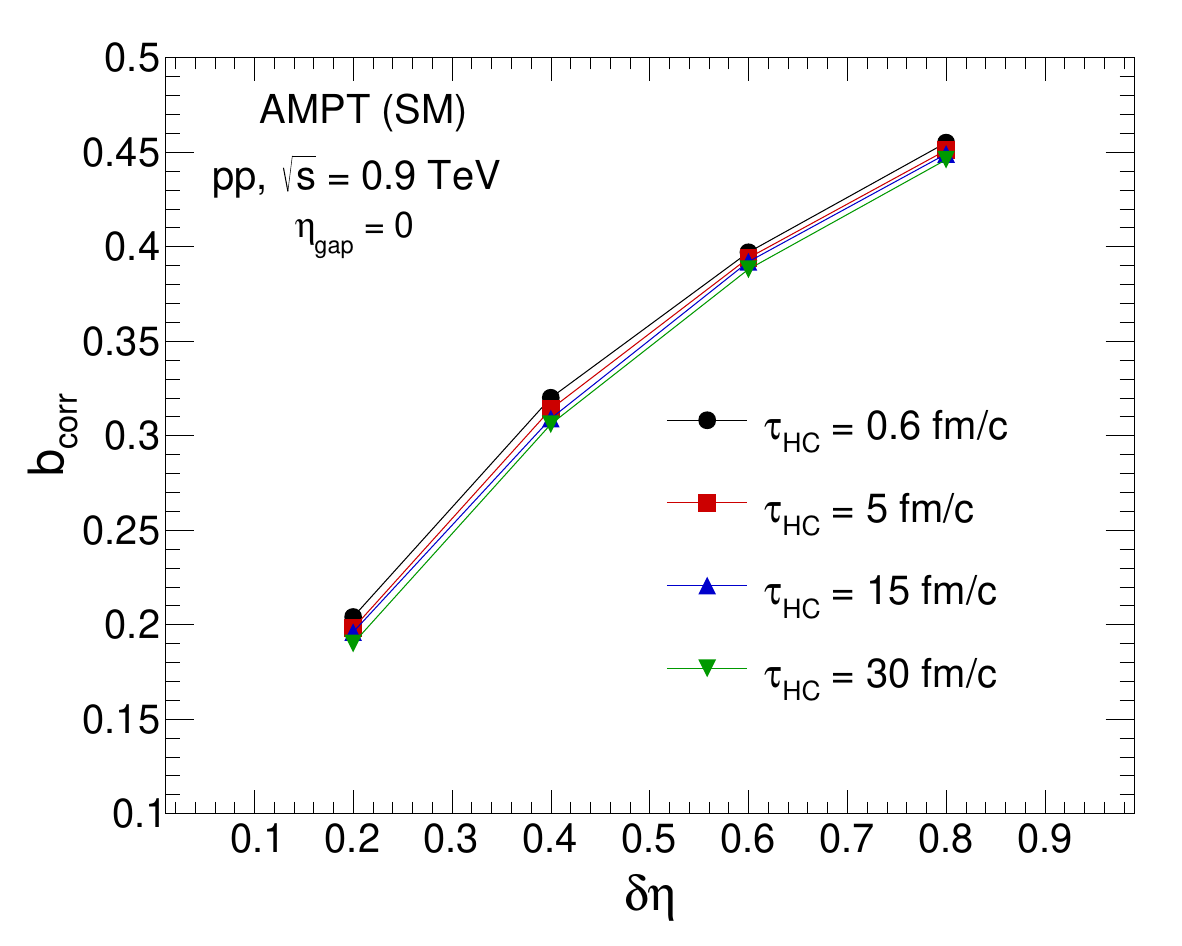} &
\includegraphics[scale=0.4]{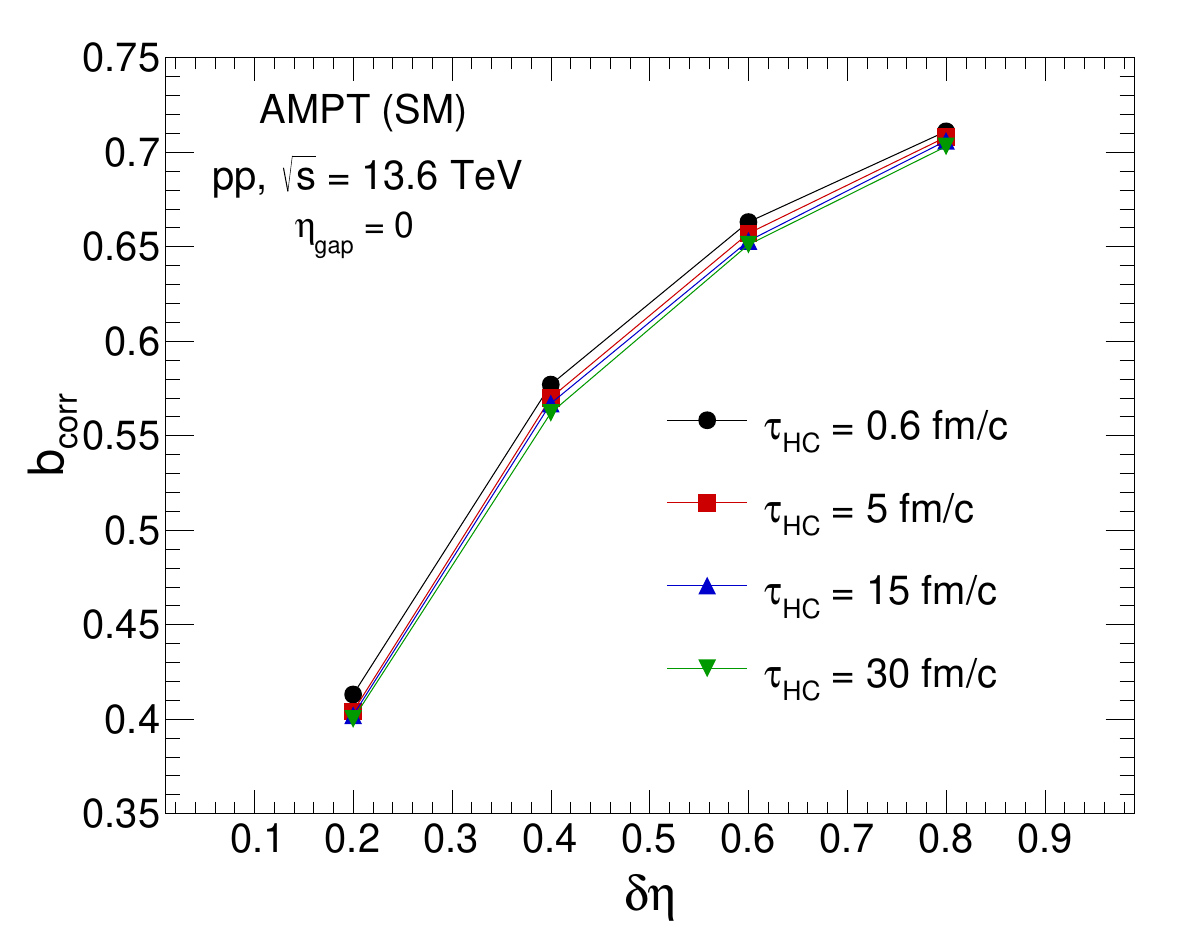} \\
(a) & (b) \\
\end{tabular}
\caption{Correlation strength $b_{\rm corr}$ as a function of the window width $\delta\eta$ for a pseudorapidity gap $\eta_{\rm gap} = 0$ in pp collisions at $\sqrt{s} = 0.9$ TeV (a) and $13.6$ TeV (b), for different hadronic cascade times $\tau_{\rm HC} = 0.6$ fm/$c$ (black), $5$ fm/$c$ (red), $15$ fm/$c$ (blue), and $30$ fm/$c$ (green). The lines joining the points are drawn to guide the eyes.}
\label{fig15}
\end{center}
\end{figure}

\begin{figure}[!h]
\begin{center}
\includegraphics[scale=0.5]{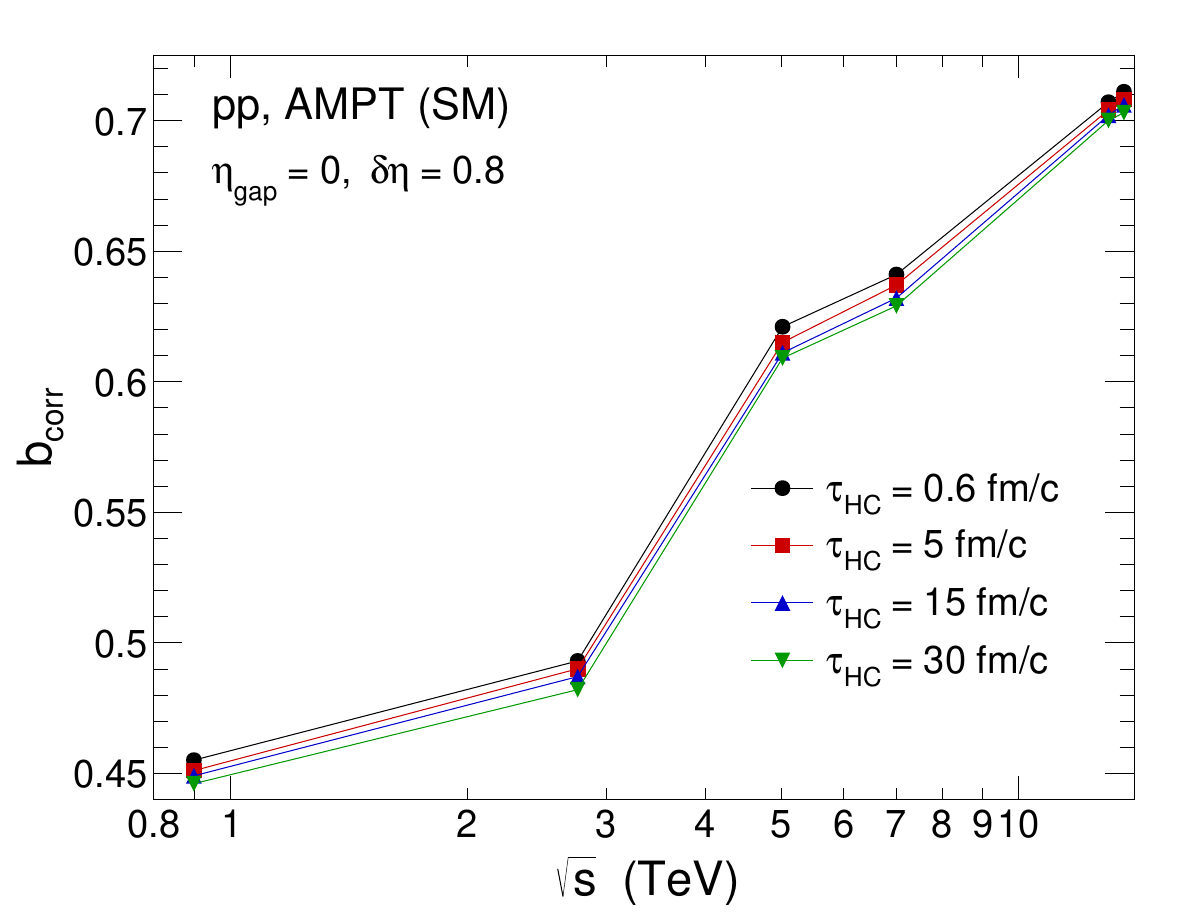}
\caption{Correlation strength $b_{\rm corr}$ as a function of the centre-of-mass energy $\sqrt{s}$ for a window width $\delta\eta = 0.8$ and a pseudorapidity gap $\eta_{\rm gap} = 0$ in pp collisions, for different hadronic cascade times $\tau_{\rm HC} = 0.6$ fm/$c$ (black), $5$ fm/$c$ (red), $15$ fm/$c$ (blue), and $30$ fm/$c$ (green).The lines joining the points are drawn to guide the eyes.}
\label{fig16}
\end{center}
\end{figure}

\section{Dependence of FB multiplicity correlations on the hadronic cascade time}
\label{sec:hadron_cascade_time}

The hadronic cascade time $\tau_{\rm HC}$ governs the duration over which the hadronic stage of the collision evolves in AMPT. To isolate its effect on the FB correlation strength, the analysis in this section is carried out exclusively with the string melting (SM) configuration, keeping the partonic stage and the initial state fixed while $\tau_{\rm HC}$ is varied. Four values are considered: $\tau_{\rm HC} = 0.6$, $5$, $15$, and $30$~fm/$c$. The results are presented for pseudorapidity-separated windows with $\eta_{\rm gap} = 0$ and for all four window widths $\delta\eta = 0.2$, $0.4$, $0.6$, and $0.8$.

\subsection{Window-width dependence at fixed energy}
\label{subsec:tauhc_deta}

Figure~\ref{fig15} shows $b_{\rm corr}$ as a function of $\delta\eta$ at $\sqrt{s} = 0.9$~TeV (panel (a)) and $13.6$~TeV (panel (b)), for the four values of $\tau_{\rm HC}$. At both energies, $b_{\rm corr}$ increases monotonically with $\delta\eta$ for every cascade time, consistent with the trend reported in Section~\ref{sec:mult_sep}. At fixed $\delta\eta$, a longer hadronic cascade leads to a systematically smaller $b_{\rm corr}$. The effect is monotonic in $\tau_{\rm HC}$, the largest values are obtained with the shortest cascade ($\tau_{\rm HC} = 0.6$~fm/$c$) and the smallest with the longest ($\tau_{\rm HC} = 30$~fm/$c$). The magnitude of the reduction is modest but resolvable. For instance, at $\sqrt{s} = 13.6$~TeV and $\delta\eta = 0.8$, $b_{\rm corr}$ falls from $0.711 \pm 0.0019$ at $\tau_{\rm HC} = 0.6$~fm/$c$ to $0.703 \pm 0.0020$ at $\tau_{\rm HC} = 30$~fm/$c$. At $\sqrt{s} = 0.9$~TeV and the same width, the corresponding values are $0.451 \pm 0.0017$ and $0.446 \pm 0.0017$. Since the statistical uncertainties on each point are of order $0.002$, the separation between the curves is well beyond the statistical precision of the simulation.

The ordering of the four curves is preserved across all four window widths and at both energies. No crossing between curves is observed, which indicates that the dependence on $\tau_{\rm HC}$ is not an artefact of a particular window configuration but a persistent feature of the model.

\subsection{Energy dependence at fixed window width}
\label{subsec:tauhc_energy}

Figure~\ref{fig16} displays $b_{\rm corr}$ as a function of the centre-of-mass energy for a fixed window width $\delta\eta = 0.8$ and $\eta_{\rm gap} = 0$, again for the four values of $\tau_{\rm HC}$. The data points shown correspond to the energies for which dedicated runs were performed: $\sqrt{s} = 0.9$, $2.76$, $5.02$, $7$, $13$, and $13.6$~TeV. At every energy, $b_{\rm corr}$ increases with $\sqrt{s}$, and at every energy the four cascade-time curves remain ordered as in Figure~\ref{fig15}, with the shortest cascade time yielding the largest $b_{\rm corr}$.

The energy dependence is considerably stronger than the cascade-time dependence. Between $\sqrt{s} = 0.9$ and $13.6$~TeV, $b_{\rm corr}$ rises from approximately $0.45$ to approximately $0.71$ for the shortest cascade time, an increase of roughly $0.26$. Over the same energy range, the spread between the $\tau_{\rm HC} = 0.6$~fm/$c$ and $\tau_{\rm HC} = 30$~fm/$c$ curves is at most about $0.01$. The cascade-time effect therefore represents a small correction on top of a dominant energy-dependent baseline.

The four curves in Figure~\ref{fig16} are approximately parallel. The spacing between adjacent cascade times does not change appreciably with energy, which suggests that the fractional reduction of $b_{\rm corr}$ due to hadronic rescattering is roughly independent of $\sqrt{s}$ over the range studied here. This is consistent with the interpretation that the hadronic cascade acts as a dilution stage whose efficiency is set by the cascade duration rather than by the collision energy.

\subsection{Interpretation}
\label{subsec:tauhc_interpretation}

The systematic reduction of $b_{\rm corr}$ with increasing $\tau_{\rm HC}$ can be understood as the effect of hadronic rescattering on the correlations established at earlier stages. The long-range component of $b_{\rm corr}$ is generated by fluctuations in the number and properties of the particle-emitting sources, which are largely determined by the initial state and the partonic stage of the collision. Extending the hadronic cascade allows more rescattering, resonance formation and decay, and particle redistribution within the acceptance, all of which tend to wash out part of the original correlation. The result is a smaller $b_{\rm corr}$ at the end of the cascade, even though the initial correlation was the same.

The modest size of the reduction indicates that the long-range correlations are largely robust against the hadronic stage in AMPT. If the correlations were dominated by short-range effects that are particularly sensitive to rescattering, the cascade-time dependence would be expected to be stronger. The observed behaviour is instead consistent with a picture in which the correlation strength is primarily imprinted at the partonic stage and only mildly modified by the subsequent hadronic evolution.

The monotonic ordering of the curves in both figures supports the interpretation that $\tau_{\rm HC}$ acts as a well-defined control parameter for the hadronic dilution. Because no crossing between curves is observed, the effect is not the result of a particular kinematic configuration or a particular energy, but a general property of the model. This makes $\tau_{\rm HC}$ a useful handle for quantifying how much of the observed correlation can be attributed to the pre-hadronic stage and how much is modified by the final-state hadronic evolution.

A more quantitative statement would require a direct decomposition of $b_{\rm corr}$ into its short- and long-range contributions as a function of $\tau_{\rm HC}$. Such a decomposition lies beyond the scope of the present work, but the results presented here provide a clear qualitative indication that the hadronic cascade acts as a mild diluter of the FB correlation strength in AMPT.

\section{Conclusions}
\label{sec:conclusions}

Forward-backward multiplicity correlations have been studied in proton--proton collisions at $\sqrt{s} = 0.9$--$13.6$~TeV with the AMPT transport model in its default (DF) and string melting (SM) configurations. The correlation coefficient $b_{\rm corr}$ was extracted for windows separated in pseudorapidity and in azimuth, and its dependence on the pseudorapidity gap, the window width, the collision energy, the transverse momentum interval, and the hadronic cascade time $\tau_{\rm HC}$ was examined. The model was first validated against the available measurements of the charged-particle pseudorapidity density and the charged-pion transverse momentum spectra. The main results are as follows.

\begin{itemize}

\item $b_{\rm corr}$ decreases gradually with increasing $\eta_{\rm gap}$ while retaining a substantial pedestal, and rises non-linearly with the window width $\delta\eta$. This non-linearity must be taken into account when comparing correlation strengths obtained under different experimental conditions.

\item The pedestal value of $b_{\rm corr}$ grows markedly with collision energy, whereas the slope of the $\eta_{\rm gap}$ dependence remains approximately constant. This indicates that the short-range contribution depends only weakly on $\sqrt{s}$, while the long-range contribution dominates in $pp$ collisions and increases appreciably with energy. Our results are consistent with the ALICE measurements \cite{alice_main}.

\item The DF configuration yields systematically larger values of $b_{\rm corr}$ than the SM configuration. The SM configuration reproduces the ALICE data \cite{alice_main} better at $0.9$ and $2.76$~TeV, whereas the DF configuration provides the closer description at $7$~TeV, suggesting that the relative weight of the partonic and hadronic stages shifts with collision energy.

\item In the $p_{\rm T}$-differential study at $\sqrt{s} = 7$~TeV, the largest deviations from the ALICE data for both configurations appear in the lowest $p_{\rm T}$ intervals, while the agreement improves at higher $p_{\rm T}$. This points to a less complete description of the short-range component in AMPT.

\item Varying the hadronic cascade time within the SM configuration over $\tau_{\rm HC} = 0.6$, $5$, $15$, and $30$~fm/$c$ produces a monotonic reduction of $b_{\rm corr}$ at all energies and for all window widths, with no crossing between the cascade-time curves. The effect is small, of the order of $0.01$ over the full cascade-time range, but statistically resolvable given the precision of the simulation. The curves are approximately parallel when plotted against $\sqrt{s}$, indicating that the fractional dilution of $b_{\rm corr}$ by hadronic rescattering is roughly independent of collision energy.

\end{itemize}

These results indicate that the FB correlation strength in $pp$ collisions, as described by AMPT, is primarily imprinted at the partonic stage and only mildly diluted by the subsequent hadronic evolution. The hadronic cascade time provides a useful control parameter for quantifying the relative contributions of the two stages.

\section*{Acknowledgement}
P. K. Haldar and D. Dhar gratefully acknowledge the financial support provided by the Department of Science and Technology and Biotechnology, Government of West Bengal, India (Sanction Memo No. 1197(Sanc.)/STBT-11012(26)/4/2024-ST SEC). Also, T. Biswas acknowledges the financial support as an Inspire fellow (No. DST/INSPIRE Fellowship/2022/IF220173) from the Department of Science and Technology, Government of India. Authors are also greatful to the Inter-University Centre for Astronomy and Astrophysics ( IUCAA), Pune, India for providing Associateship programme under which a part of this work was carried out.

\section*{Data availability}

Data will be made available on request.

\section*{Declaration of competing interest}
The authors declare that they have no known competing financial interests or personal relationships that could have appeared to influence the work reported in this paper.

\end{document}